\documentclass[a4paper,11pt]{article}
\pdfoutput=1 
\usepackage{jcappub} 
\usepackage[T1]{fontenc} 
\usepackage{graphicx}
\usepackage{subfigure}

\title{Cosmic Acceleration from Topological Considerations II: Fiber Bundles}

\author[a]{Maribel Hern\'andez M\'arquez,}
\author[b]{Tonatiuh Matos Chassin and}
\author[c]{Petra Wiederhold}

\affiliation[a]{Facultad de Ciencias F\'isico-Matem\'aticas, Ciudad Universitaria, Benem\'erita Universidad Aut\'onoma de Puebla, Av. San Claudio SN, Col. San Manuel, Puebla, M\'exico.}
\affiliation[b]{Departamenteo de F\'isica, Centro de Investigaci\'on y de Estudios Avanzados del IPN, A.P. 14-740, 07000, Ciudad de M\'exico, M\'exico}
\affiliation[c]{Departamenteo de Control Autom\'atico, Centro de Investigaci\'on y de Estudios Avanzados del IPN, A.P. 14-740, 07000, Ciudad de M\'exico, M\'exico}

\emailAdd{marihm111@gmail.com}
\emailAdd{tmatos@fis.cinvestav.mx}

\abstract{In this work we study an alternative topological model for explaining the observed acceleration of space-time, we answer the question of whether this acceleration could be a consequence of the topology of the universe. For doing that, we propose that the whole universe is composed of a four dimensional base space, which represents space-time, endowed with a fiber forming a principal fiber bundle. We analyze this hypothesis for a homogeneous and isotropic four dimensional space-time and show that the effect of the fiber onto the base space is that the space-time accelerates depending on the group of the fiber, even in an oscillatory way, resembling the behavior of the universe according to recent observations. We conclude that there is the possibility of the accelerating behavior of the universe being due to its whole topology instead of an exotic kind of matter.}
\keywords{dark energy, principal fiber bundle}
\begin{document}
\maketitle
\flushbottom

\section{Introduction}
\label{sec:intro}

Doubtless the beginning of this century has been very interesting for science. We have discovered many new aspects of nature, but many fundamental questions about the behavior of matter are still open. One of the most important questions that remains unanswered is, doubtless, the nature of the dark sector of the universe. We now know that the universe contains at least two components that are almost disconnected from the rest of the matter, dark energy, which is gravitationally repulsive and dark matter, attractive. Untill now, the contact with these two components is only through gravitational interaction. There are no signs in any other way. The favorite candidates for dark matter are weak interacting particles coming from a hypothetical supersymmetry, called WIMPS, and nowadays with better perspectives an ultralight boson particle called the Scalar Field Dark Matter (see for example \cite{L9}, \cite{Ostriker}). For the dark energy sector there are many candidates like the quintessence, alternative theories of gravity, the cosmological constant, among others. To complicate the situation even more, all these favorite candidates of dark energy seem to be in big tension with recent data, because new observations of the dark energy equation of state of the universe seems to vary for different redshifts, even going beyond of minus one, which could imply the existence of a dark energy component that violates the weak energy condition (see for example \cite{Veto}). These results give rise to new hypotheses of the nature of dark energy. 

In this paper we pretend to give a new explanation of the accelerating expansion of the universe. For doing that, we continue the idea started in \cite{GarciaAspeitia:2009zz}, considering that the cosmic acceleration could be due to the topology of the whole universe, now using principal fiber bundles. The main goal here is not yet to fit dates with some candidate or to compare this candidate with observations, but to present an idea that could be an alternative to the candidates we have so far. The idea is very simple, we start from the fiber bundle formulation of quantum field theory and using the generalization of this formulation to a curved space-time \cite{Trautman:1970cy}, we explore the evolution of a homogeneous and isotropic universe, probing some groups, compact and non-compact for the fiber and show that the topological hypothesis that the universe is a curved space-time with an internal group as fibration can be an alternative to explain the accelerating expansion of the universe. We show that the base space expands with an acceleration that resembles the behavior of the real universe, in concordance with the most recent observations (see for example \cite{Veto}). We compare the behavior using  different groups and observe that this behavior strongly depends on the structure group of the fiber. In other words, we show that the observed acceleration of the space-time could be a consequence of the global topology of the whole universe composed of a base space with a fiber.

This paper is organized as follows: in section \ref{section2} we set the mathematics we need for developing our hypothesis. In section \ref{section3} we write the field equations for a one and a two dimensional Lie group.  And in \ref{section4}  and \ref{section5} we analyze the effects on the base space when we consider $SU(1,1)$ and $SU(2)$ as fibers, in order to study a non-compact and a compact fiber. Section \ref{sec:solutions} is devoted to the solutions of the field equations of section \ref{section4} and \ref{section5}. Finally in section \ref{sec:conclusions} we give our conclusions.  

\section{Mathematical preliminaries}
\label{section2}

We know that matter has not necessarily driven the universe expansion. In order to see that, we write the Friedmann equation in its simple form,
\begin{eqnarray}
\frac{1}{2}\dot a^2 -\frac{\kappa}{6}\rho a^2-\frac{\Lambda}{6}a^2=-\frac{k}{2},
\end{eqnarray}
where $a$ is the scale factor of the universe, $\rho$ is the matter density and $k$ is the curvature parameter. This equation is a dynamical system with potential $V= -\frac{\kappa}{6}\rho a^2-\frac{\Lambda}{6}a^2$ and total energy $k/2$. Here the cosmological constant $\Lambda$ comes from the geometry of space-time, and we see that even if there is no matter in the universe ($\rho=0$), space-time will expand, provided that the geometrical part $\Lambda >0$. Therefore we have strong reasons to think that the topology of space-time could be the cause of the universe expanding with some acceleration, and not necessarily a strange kind of matter.

In order to see that, we start with the following hypothesis, we follow the fiber bundle formulation of quantum field theory in curved space-time \cite{Trautman:1970cy},  \cite{Cho}, \cite{kerner}.
Thus, we consider a principal fiber bundle $P$ with projection $\pi$ and the diffeomorphism $\varphi$ called the trivialization, $\varphi:P \rightarrow U \times G$, endowed with a connection whose fiber is a Lie group $G$ and base a four-dimensional pseudo-Riemannian manifold $U\subset B$. These assumptions define a metric in P because the connection separates the tangent space of P
\begin{equation}
    T(P)=V(P)\oplus H(P)
\end{equation}
into their vertical and horizontal subspaces.

Since $T(P)$ has dimension $4+d$ ($d$ is the dimension of the group), and $V(P)$ has dimension $d$, then $H(P)$ has dimension 4. Now let $\left\{\hat{e}_{\alpha}\right\}$ the basis for the horizontal space, $\left\{\hat{e}_{i}\right\}$ the bases for the vertical space and $\left\{\hat{\omega}^{\beta}\right\}$, $\left\{\hat{\omega}^j\right\}$ their corresponding dual basis, where $\alpha, \beta=1,...,4$ and $i,j=5,...,n+d$. With the following assumptions, we can define a unique metric in $P$ compatible with the metric of the base space and the metric of the group (see \cite{Cho})
\begin{eqnarray}
\hat{g}(\hat{e}_i,\hat{e}_j)=\tilde{g}(d\varphi(\hat{e}_i),d\varphi(\hat{e}_j))\nonumber\\
\hat{g}(\hat{e}_{\alpha},\hat{e}_{\beta})=0\nonumber\\
\hat{g}(\hat{e}_{\alpha},\hat{e}_{\beta})=g(d\pi(\hat{e}_{\alpha}),d\pi(\hat{e}_{\beta}))
\end{eqnarray}
where $\tilde{g}$, $g$ and $\hat{g}$ are the metrics on $G$, the base space $B$ and $P$ respectively.

Now we will write $\hat{g}$ in local coordinates. The projection of the horizontal space is non-zero and forms a basis of the tangent space of $U$, i.e.
\begin{equation}
d\pi(\hat{e}_{\alpha})=e_{\alpha}.
\end{equation}
On the other hand, the projection of vertical space is zero:
\begin{equation}
d\pi(\hat{e}_{i})=0.
\end{equation}

Thus, if we project the vectors  $\left\{ \hat{e}_{\alpha},\hat{e}_i\right\}$ to the tangent space $U \times G$ through the trivialization $\varphi$, we obtain, \begin{eqnarray}
d\varphi(\hat{e}_{\alpha})=B^{\beta}_{\alpha}e_{\alpha}-A^{j}_{\alpha}e_j\nonumber\\
d\varphi(\hat{e}_i)=C^{\beta}_i e_{\beta}+D^j_i e_j,
\end{eqnarray}
where $\left\{e_j\right\}$ is a left invariant basis of the tangent space of $G$ such  that  $\left\{e_{\alpha},e_j\right\}$ is a basis of $T(U \times G)$, and $A_\alpha^j$, $B^{\beta}_{\alpha}$, $C^{\beta}_i$ and $D^j_i$ are arbitrary coefficients. 

 Now, we consider the projection from $U\times G$ into $U$,   $\pi_1: U\times G \rightarrow U, (x,a)\mapsto x$ such that
\begin{equation}
\label{proyeccion}
\pi=\pi_1\circ\varphi
\end{equation}
So, we have
\begin{eqnarray}
d\pi(\hat{e}_{\alpha})=d\pi_1\circ d\varphi(\hat{e}_{\alpha})=B^{\beta}_{\alpha}e_\beta=e_{\alpha}\nonumber\\
d\pi(\hat{e}_i)=d\pi_1\circ d\varphi(\hat{e}_i)=C^{\beta}_i e_{\beta}=0,
\end{eqnarray}
$i.e.$,
$B^{\beta}_{\alpha}=\delta^{\beta}_{\alpha}$ y $C^{\beta}_i=0$. The set $d\varphi(\hat{e}_i)=D^j_i e_j$ is a  basis of $T(G)$ and we can rewrite it as $D^{j}_i e_j\rightarrow e_i$. So we have 
\begin{eqnarray}
\label{basis}
d\varphi(\hat{e}_{\alpha})=e_{\alpha}-A^{i}_{\alpha}e_i\nonumber\\
d\varphi(\hat{e}_i)=e_i
\end{eqnarray}
We can easily find the dual base of (\ref{basis}), we arrive at
\begin{equation}
\bar{e}_A=\left \{
      \begin{array}{rcl}
          e_{\alpha}-A^{j}_{\alpha}e_j \\
          e_j \\ 
      \end{array}
   \right.
   \end{equation}
   \begin{equation}
   \bar{\omega}^A=\left \{
      \begin{array}{rcl}
          \omega^{\alpha}\\
          \omega^i+A^i_{\alpha}\omega^{\alpha}
      \end{array}
   \right.
   \end{equation}
where $\left\{\omega^{\beta}\right\}$ is the dual of $\left\{e_{\alpha}\right\}$, $\left\{\omega^{m}\right\}$ is the dual of $\left\{e_{n}\right\}$ and $A=1,2,..,4+d$.
And finally with this basis we can write the local metric $\bar{g}$ as
\begin{equation}
\label{metricalocal}
\bar{g}=g_{\alpha\beta}\omega^{\alpha}\otimes\omega^{\beta}+I_{ij}(\omega^i+A^i_{\alpha}\omega^{\alpha})\otimes(\omega^j+A^{j}_{\beta}\omega^{\beta})
\end{equation}
It can be shown that  
\begin{equation}
\hat{g}=\varphi^*\bar{g}
\end{equation}
where $\varphi^*$ is the pullback of $\varphi$ and  $A^{i}_{\beta}\omega^{\beta}t_i$ is the projection of the one-form of connection to $U\subset B$ and they are interpreted as the Yang-Mills-gauge fields \cite{Trautman:1970cy}.

In what follows, we shall use Greek indices $\alpha,\,\beta ...=1,..,4$ for the four dimensional space-time $B$, Latin indices $i,\,j ...=5,..,4+d$ for the group $G$ and capital indices $A,\,B ...=1,..,4+d $ for the whole fiber $P$.

\section{The Field Equations}
\label{section3}

We will focus on the cosmology of this model. We start with a homogeneous and isotropic four dimensional space-time, $i.e.$, with a homogeneous and isotropic base space $B$. For doing so, we set the Friedmann-Lemaitre-Robertson-Walker (FLRW) metric in the four dimensional base space and use (\ref{metricalocal}) with different groups to write the field equations. The first question we want to answer here is whether the hidden internal symmetry could be observable, in other words, we want to see the effects of the fiber onto space-time and finally we want to see whether these effects are related with the dark energy.  

To explore the idea that the universe is a principal fiber bundle, we first consider a one and two-dimensional Lie group and in section (\ref{section4}) two specific three-dimensional Lie groups, SU(1,1) and SU(2). The last two groups are very different, the first one is non-compact and the second one is compact. We choose these groups only to compare the different effects of the fibre onto space-time. In the global space-time the Yang-Mills fields can be perfectly neglected, thus we set  $A^{i}_{\beta}\omega^{\beta}=0 $ in metric (\ref{metricalocal}), thus, the local metric of $P$ is simply the metric of the base space plus the metric of the Lie group. We can do this because the region $U\subset B$ that we choose has to be  big enough  to ignore the interactions between particles.

An important difference of previous works \cite{Cho}, \cite{kerner} is that we use a left-invariant metric on the fiber, we do not restrict our work to a bi-invariant metric which has the following form
\begin{equation}
    g_{ij}w^i \otimes w^j
\end{equation}
where $\left\{w^i\right\}$ is a basis for the left-invariant 1-forms and 
\begin{equation}
    g_{ij}=-c^k_{im}c^m_{jk}
\end{equation}
where $c^i_{jk}$ are the structure constants of the Lie algebra of the group G.
This is because in the case of a one dimensional Lie group and two-dimensional Lie groups, we can't define a bi-invariant metric, because in the first case the only structure constant is equal to zero and in the case of an Abelian two dimensional Lie group, all the structure constants are zero. While for a non-Abelian two dimensional Lie group the matrix $(g_{ij})$ is singular. So if we want to explore the idea that our universe is a principal fiber bundle with a fiber of one and two dimensions, we have to use a left invariant metric that can be constructed as follows.

 Let $G$ be a Lie Group and let $\left\{\omega^5,..,\omega^{4+d}\right\}$  be a basis for the left-invariant 1 forms; if $(a_{ij})$ is any (constant) non-singular symmetric $n\times n$ matrix, then
\begin{equation}
\label{leftinvariantmetric}
a_{ij}w^i \otimes w^j,
\end{equation}
is a metric tensor on $G$, which is a left-invariant metric \cite{Torres}.

In what follows, we specialize these results for the one and two-dimensional Lie groups.

\subsection{One dimensional Lie group}

The metric is
\begin{eqnarray}
\bar{g}= dt \otimes dt-\frac{a(t)^2}{{1-kr}}dr \otimes  dr-a(t)^2r^2(d\theta\otimes d\theta-\sin\theta^2 \phi\otimes d \phi)
- b(t)^2(w^5 \otimes w^5) 
\end{eqnarray}
where $w^5$ is a left-invariant 1-form whose exterior derivative is zero. Let us introduce 
\begin{eqnarray}
\label{1formone}
\theta^1\equiv& dt\nonumber\\
\theta^2\equiv&\frac{a(t)}{1-kr^2}dr\nonumber\\
\theta^3\equiv &a(t)rd\theta\nonumber\\
\theta^4\equiv &a(t)r\sin\theta d\phi\nonumber\\
\theta^5\equiv &b(t)w^5,
\end{eqnarray}
the previous 1-forms are the dual basis of an orthonormal basis, such that, $\bar{g}=\eta_{AB}\theta^A\otimes \theta^B$, with
\begin{equation}
(\hat{\eta}_{AB})=\text{diag}\left\{1,-1,-1,-1,-1,\right\}
\end{equation}
that represents a rigid basis. 
From the second Cartan structural equations,
\begin{equation}
\label{cartan}
\mathcal{R}_{AB}=d\Gamma_{AB}+\Gamma_{AC}\wedge\Gamma^{C}_B=\frac{1}{2}R^A_{BCD}\theta^C\wedge\theta^D    
\end{equation}
we can get the components of the Riemann tensor. For that, we need to know the connection 1-forms given by
\begin{equation}
\label{connection1form}
    \Gamma_{AB}\equiv g_{AC}\Gamma^C_B=\Gamma_{ABC}\theta^C,
\end{equation}
being $\Gamma^A_{CB}$ the Ricci rotation coefficients.  We can obtain these from the first Cartan structural equations, owing to the fact that the torsion of the Riemannian connection is equal to zero and that the exterior product of 1-forms is skew-symmetric, we have
\begin{equation}
\label{dtheta}
    d\theta^A=\Gamma^A_{BC}\theta^B\wedge\theta^C=\Gamma^A_{[BC]}\theta^B\wedge\theta^C
\end{equation}
and 
\begin{equation}
\label{riccicoef}
    \Gamma_{CBA}=\Gamma_{C[BA]}-\Gamma_{B[CA]}-\Gamma_{A[CB]}.
\end{equation}

Taking the exterior derivative of (\ref{1formone}) and according to (\ref{dtheta}),
\begin{eqnarray}
\label{antisi}
\Gamma^2_{[12]}=\Gamma^3_{[13]}=\Gamma^4_{[14}]=&\frac{1}{2}\frac{\dot{a}}{a}\nonumber\\
\Gamma^3_{[23]}=\Gamma^4_{[24]}=&\frac{1}{2}\frac{\sqrt{1-kr^2}}{ar}\nonumber\\
\Gamma^4_{[34]}=&\frac{1}{2}\frac{\cot\theta}{ar}\nonumber\\
\Gamma^5_{[15]}=&\frac{\dot{b}}{4b}.
\end{eqnarray}
Replacing (\ref{antisi}) in (\ref{riccicoef}) and then with (\ref{connection1form}) and (\ref{cartan}), we can find the components of the Riemann tensor and then the components of the Einstein's tensor. Using a  diagonal energy-momentum tensor given by
\begin{equation}
\label{tensorenergia}
T_{AB}=\text{diag}(\rho(t),p(t),p(t),p(t),0),
\end{equation}
the Einstein's equations read
\begin{subequations}
\begin{eqnarray}
\frac{\dot{a}\dot{b}}{ab}+\frac{k}{a^2}+\frac{\dot{a}^2}{a^2}=\kappa\frac{\rho}{3}\\
 2\frac{\ddot{a}}{a}+\frac{\ddot{b}}{b}+2\frac{\dot{a}\dot{b}}{ab}+\frac{k}{a^2}+\frac{\dot{a}^2}{a^2}=-\kappa p\\
\label{factorscaleone}
    \frac{\ddot{a}}{a}+\frac{\dot{a}^2}{a^2}=0.
\end{eqnarray}
\end{subequations}

From (\ref{factorscaleone}), one finds
\begin{equation}
    a(t)=\pm\sqrt{c_1t+c_2}.
\end{equation}
The physical solution is that with the positive sign, so if we set $a(t=0)=0$ the solution is $a(t)=\sqrt{c_1t}$ and 
\begin{equation}
    \ddot{a}=-\frac{1}{4}\frac{\sqrt{c_1}}{t^{3/2}},
\end{equation}
that is to say, when the fiber is the group $U(1)$ the scale factor always slows down.

Also, we can combine the previous equations to obtain
\begin{equation}
\label{combinado}
    3\frac{\ddot{a}}{a}+\frac{2}{3}\kappa\rho+\frac{\ddot{b}}{b}-\frac{k}{a^2}=-\kappa p.
\end{equation}
We can replace $a(t)$ and $\ddot{a}$ in (\ref{combinado}) to obtain an equation for $b(t)$ that cannot be integrated directly. One could solve it numerically but we consider it unnecessary for the purpose of this paper.

\subsection{Fibers of two dimensional Lie groups}

  Let $\left\lbrace w^5, w^6 \right\rbrace$ be a basis for the left-invariant 1-forms of a two dimensional Lie group. According to (\ref{leftinvariantmetric}), a left invariant metric for a two dimensional Lie group is
  \begin{equation}
      b(t)^2w^5\otimes w^5+c(t)^2 w^6\otimes w^6.
      \end{equation}
  Then the metric (\ref{metricalocal}) transforms into
\begin{eqnarray}
    \bar{g}= dt \otimes dt-\frac{a(t)^2}{{1-kr}}dr \otimes  dr-a(t)^2r^2(d\theta\otimes d\theta-\sin\theta^2 d \phi\otimes d \phi)\nonumber\\+ b^2(t)(w^5 \otimes w^5)+c^2(t)(w^6 \otimes w^6).
\end{eqnarray}
We can define 
\begin{eqnarray}
\theta^1\equiv&dt\nonumber\\
\theta^2\equiv&\frac{a(t)}{\sqrt{1-kr^2}}dr\nonumber\\
\theta^3\equiv&a(t)rd\theta\nonumber\\
\theta^4\equiv&a(t)r\sin\theta d\phi\nonumber\\
\theta^5\equiv&b(t)\omega^5\nonumber\\
\theta^6\equiv&c(t)\omega^6,
\end{eqnarray}
the previous 1-forms are the dual basis of an orthonormal basis, such that, $\bar{g}=\eta_{AB}\theta^A\otimes \theta^B$, with
\begin{equation}
(\eta_{AB})=\text{diag}\left\{1,-1,-1,-1,-1,-1\right\}.
\end{equation}

Also, we know that if {$w^i,...,w^{n+d}$} is a basis for the 1-forms on $G$, then 
\begin{equation}
\label{maurer-cartan}
    dw^i=-\frac{1}{2}C^i_{jk}w^i\wedge w^k,
\end{equation}
where $C^i_{jk}$ are the structure constants with respect to the basis $\left\lbrace e_i,...,e_{n+d}\right\rbrace$.

We know that for an Abelian Lie group of two dimensions, the structure constants are all zero and that for a non-Abelian  we can conveniently choose a basis of the Lie algebra $\left\lbrace e_5,e_6 \right\rbrace$ such that the only structure constants different from zero are
\begin{equation}
\label{nonabel}
 c^6_{56}=1,\hspace{1cm} c^6_{65}=-1.  
 \end{equation}

So when we have an Abelian Lie group of two dimensions using (\ref{maurer-cartan}), we obtain
\begin{equation}
    d\theta^5=\frac{\dot{b}}{b}\theta^1\wedge\theta^5\hspace{1cm}d\theta^6=\frac{\dot{c}}{c}\theta^1\wedge\theta^6,
\end{equation}
then, 
\begin{equation}
    \Gamma^5_{[15]}=\frac{1}{2}\frac{\dot{b}}{b}, \hspace{1cm}\Gamma^6_{[16]}=\frac{1}{2}\frac{\dot{c}}{c},
\end{equation}
and for a non-abelian, using (\ref{nonabel}),
\begin{equation}
    d\theta^5=\frac{\dot{b}}{b}\theta^1\wedge\theta^5\hspace{1cm}
d\theta^6=\frac{\dot{c}}{c}\theta^1\wedge\theta^6-\frac{1}{b}\theta^5\wedge\theta^6
\end{equation}
then, 
\begin{equation}
   \Gamma^5_{[15]}=\frac{1}{2}\frac{\dot{b}}{b} \hspace{1cm}\Gamma^6_{[16]}=\frac{1}{2}\frac{\dot{c}}{c},\hspace{1cm}\Gamma^6_{[56]}=-\frac{1}{2b}.
\end{equation}
With the previous expressions and from the second Cartan structural equations,  we can get the components of the Einstein's tensor for an Abelian  and a non-Abelian Lie group of two dimensions. To obtain the Einstein's equations, we consider a diagonal energy momentum tensor
\begin{equation}
\label{tensorenergia6}
T_{AB}=\text{diag}(\rho(t),p(t),p(t),p(t),0,0).
\end{equation}
Then, the Einstein's equations for an Abelian Lie group are
%
\begin{subequations}
\begin{eqnarray}
\label{abel1}
3\left(\frac{k}{a^2}+\frac{\dot{a}^2}{a^2}\right)+3\frac{\dot{a}}{a}\left(\frac{\dot{c}}{c}+\frac{\dot{b}}{b}\right)+\frac{\dot{b}\dot{c}}{bc}&=&\kappa\rho\\
\label{abel2}
    2\frac{\Ddot{a}}{a}+\frac{\Ddot{b}}{b}+\frac{\Ddot{c}}{c}+2\frac{\dot{a}}{a}\left(\frac{\dot{b}}{b}+\frac{\dot{c}}{c}\right)+\frac{\dot{b}\dot{c}}{bc}+\frac{k}{a^2}+\frac{\dot{a}^2}{a^2}&=&-\kappa p\\
\label{abel3}
    3\frac{\Ddot{a}}{a}+\frac{\Ddot{c}}{c}+3\left(\frac{k}{a^2}+\frac{\dot{a}^2}{a^2}\right)+\frac{3\dot{a}\dot{c}}{ac}&=&0\\
\label{abel4}
      3\frac{\Ddot{a}}{a}+\frac{\Ddot{b}}{b}+3\left(\frac{k}{a^3}+\frac{\dot{a}^2}{a^2}\right)+3\frac{\dot{a}\dot{b}}{ab}&=&0.
\end{eqnarray}
\end{subequations}

And from the conservation of the energy momentum tensor
\begin{equation}
\label{conservation}
    \frac{\dot{\rho}}{\rho}+3\frac{\dot{a}}{a}(1+w)+\left(\frac{\dot{b}}{b}+\frac{\dot{c}}{c}\right)=0.
\end{equation}
If we further assume that the perfect fluid obeys a barotropic equation of state $p=w\rho$, this equation can be integrated to obtain
\begin{equation}
\label{density2}
\rho=\frac{\rho_0b_0c_0}{a^{3(1+w)}bc},
\end{equation}
where the  zero denotes the present value.

Meanwhile, the Einstein's equations for a non-Abelian two dimensional Lie group read as
\begin{subequations}
\begin{eqnarray}
\label{non1}
-\frac{\dot{b}}{b^2}+\frac{\dot{c}}{cb}&=&0\\
\label{nonabel1}
3\left(\frac{\dot{a}^2}{a^2}+\frac{k}{a^2}\right)+\frac{\dot{a}}{a}\left(\frac{\dot{b}}{b}+\frac{\dot{c}}{c}\right)-\frac{1}{b^2}+\frac{\dot{b}\dot{c}}{bc} &=&\kappa\rho\\
\label{nonabel2}
    2\frac{\Ddot{a}}{a}+\frac{\Ddot{b}}{b}+\frac{\Ddot{c}}{c}+\frac{k}{a^2}+\frac{\dot{a}^2}{a^2}+2\left(\frac{\dot{a}}{a}+\frac{\dot{c}}{c}\right)-\frac{1}{b^2}+\frac{\dot{b}\dot{c}}{bc}&=&-\kappa  p\\
    \label{nonabel3}
    3\frac{\Ddot{a}}{a}+\frac{\Ddot{c}}{c}+3\left(\frac{k}{a^2}+\frac{\dot{a}^2}{a^2}\right)+3\frac{\dot{a}\dot{c}}{ac}&=&0\\
\label{nonabel4}
     3\frac{\Ddot{a}}{a}+\frac{\Ddot{b}}{b}+3\left(\frac{k}{a^2}+\frac{\dot{a}^2}{a^2}\right)+3\frac{\dot{a}\dot{b}}{ab}&=&0.
\end{eqnarray}
\end{subequations}
In this case, the  density is also given by (\ref{density2}).

Equation (\ref{non1}) implies that 
\begin{equation}
\label{igualdad}
  c(t)=c_1b(t),     
\end{equation}
where $c_1=$ constant.
Using (\ref{igualdad}), the previous equations are simplified and the equations (\ref{nonabel3}) and (\ref{nonabel4}) are the same.
We solve the field equations with $k=0$, for that we rewrite them in terms of the density parameter; i.e.,
\begin{equation}
\label{densityparameter}
\Omega =\frac{\rho}{\rho_c},    
\end{equation}
where $\rho_c$ is the critical density,
\begin{equation}
    \rho_c=\frac{3H^2}{8\pi G}.
\end{equation}
We consider the total density as
\begin{equation}
\label{densidadtotal}
   \rho=\rho_m+\rho_{\gamma}=\frac{\rho_{0m}b_0c_0}{a^3bc}+\frac{\rho_{0\gamma}b_0c_0}{a^4bc},
\end{equation}
where $\rho_m$ is the density of the baryonic matter plus dark matter, $\rho_{\gamma}$ is the density of radiation and $\rho_{0m}$, $\rho_{0\gamma}$ are the corresponding present values. 
So 
\begin{equation}
\label{k4densidad}
    \kappa\rho=\frac{8\pi G}{3H_0^2}\rho H_0^2=\Omega_{0m}H_0^2\frac{b_0c_0}{a^3bc}+\Omega_{0\gamma}H_0^2\frac{b_0c_0}{a^4bc},
\end{equation}
where $H_0$ is the present value of the Hubble's constant.
We replace (\ref{k4densidad}) in the field equations of an Abelian and non-Abelian two dimensional Lie group and we rewrite them in terms of a dimensionless variable $X=H_0 t$.
With the aforementioned procedure, we start solving the equations for an Abelian Lie group when $b(t)=c(t)$ numerically. In this case we find from (\ref{abel1})-(\ref{abel4}),
\begin{eqnarray}
    a''&=&\frac{\Omega_{0\gamma}b_0^2}{3a^3b^2}+\frac{\Omega_{0m}b_0^2}{4a^2b^2}+4\frac{a'^2}{a}\mp 2a'\left(6\frac{a'^2}{a^2}+\frac{\Omega_{0m}b_0^2}{a^3b^2}+\frac{\Omega_{0\gamma}b_0^2}{a^4b^2}\right)^\frac{1}{2}\nonumber\\
b'&=&-3\frac{a'}{a}b\pm b\left(6\frac{a'^2}{a^2}+\frac{\Omega_{0m}b_0^2}{a^3b^2}+\frac{\Omega_{0 \gamma} b_0^2}{a^4 b^2}\right)^\frac{1}{2},
\end{eqnarray}
where the primes denote derivatives with respect to $X$.
We solve these equations numerically as a dynamical system, considering the following initial conditions, 
$a_i=.001$, $0<a_i'$ and $b_i\neq 0$, where $i$ denotes the initial value.
It is easy to see that we have two cases, the first one is when we consider $a''$ with the negative sign and $b'$ with the positive one. Due to the fact that the previous equations are true for any value of $t$, it is possible to find initial conditions that give rise to an initial acceleration $a_i''<0$ and others with $a_i''>0$. We found that if   $a_i''<0$,  the scale factor grows and $a''$ is always negative and its magnitude  tends to zero.

On the other hand, when we consider $a''$ with the positive sign and $b'$ with the negative, we can see from the equation for $a''$ that $a_i''$ is always positive, no matter the value of the initial conditions. But, according to observations, we are interested in solutions with an initial decelerating expansion that in some point of the evolution of the universe becomes accelerating. So this solution is not physical.\\
When $b(t)\neq c(t)$ from (\ref{abel1}) to (\ref{abel4}), we can find the following expressions:
\begin{eqnarray}
a''&=&\frac{\Omega_{0\gamma}b_0c_0}{12a^2bc}-\frac{1}{8}\frac{a'c'}{a}-\frac{5}{4}\frac{a'^2}{a}
+\left(\frac{\Omega_{0\gamma}b_0c_0}{a^3bc}+\frac{\Omega_{0m}b_0c_0}{a^2bc}-3\frac{a'^2}{a}-\frac{3a'c'}{c}\right)\left(\frac{-a'c+c'a}{12a'c+4ac'}\right)\nonumber\\
\\
b'&=&\left(\frac{\Omega_{0\gamma}b_0c_0}{a^4bc}+\frac{\Omega_{0m}b_0c_0}{a^3bc}-3\frac{a'^2}{a^2}-3\frac{a'c'}{ac}\right)\left(\frac{abc}{3a'c+ac'}\right)
\\ \nonumber\\
c''&=&-\frac{\Omega_{0\gamma}b_0c_0}{4a^4b}+\frac{c'^2}{2c}-\frac{9}{4}\frac{a'c'}{a}+\frac{3}{4}\frac{a'^2}{a^2}c-\frac{c'^2}{c}
\left(\frac{\Omega_{0\gamma}b_0c_0}{a^4b}+\frac{\Omega_{0m}b_0c_0}{a^3b}-3\frac{a'^2}{a^2}c-3\frac{a'c'}{a}\right)\left(\frac{3a'c-3c'a}{12a'c+4ac'}\right).\nonumber\\
\end{eqnarray}
Considering the following initial conditions,  $a_i=.001$, $a'_i>0$, $b_i\neq 0$, $c_i\neq 0$, $c'_i\neq 0$, we solved the previous equations like a dynamical system and we find that the solutions only reproduce a decelerating universe, that is to say , there are only solutions with $a_i''<0$ and $a''<0$.

For a two dimensional non-Abelian Lie group, the only case to consider is $c(t)=c_1b(t)$. From (\ref{nonabel1})-(\ref{nonabel4}),
\begin{eqnarray}
\label{nonabelace}
a''&=&\frac{\Omega_{0\gamma}b_0}{a^3b}+\frac{\Omega_{0m}b_{0}}{4a^2b}\mp2a'\left(6\frac{a'^2}{a^2}+\frac{1}{bH_0^2}+\frac{\Omega_{0m}b_{0}}{a^3b}+\frac{\Omega_{0\gamma}b_0}{a^4b}\right)^\frac{1}{2}\nonumber\\
b'&=&-6\frac{a'}{a}b\pm 2b\left(6\frac{a'^2}{a^2}+\frac{1}{bH_0^2}+\frac{\Omega_{0m}b_0}{a^3b}+\frac{\Omega_{0\gamma}b_0}{a^4b}\right)^\frac{1}{2}.
\end{eqnarray}

There are again 2 cases. The first one is when one considers in (\ref{nonabelace}) the negative sign for $a''$ and the positive  for $b'$. In this case, depending on the values of the initial conditions, $a_i''$ can be positive or negative. The physical solution is when $a_i''<0$. When this happens, it can be found that $a''$ is negative and the magnitude of $a''$ tends to zero. Therefore it doesn't reproduce the observed universe.
The second case is when one considers in (\ref{nonabelace}) the positive sign, because this is true for any value of $t$; $a_i''$ is always positive no matter the value of the initial conditions, again this case doesn't mimic our universe.
The most important result of this section is that a principal fiber bundle with a fiber of two dimensions doesn't reproduce the evolution of the scale factor. The equations and the results found in this section are valid for any two dimensional Abelian Lie-group  and non-Abelian. So, a fiber like $U(1)\times U(1)$ doesn't reproduce our observed universe.

\section{Fiber bundle on the SU(1,1) group}
\label{section4}

We consider again for the base space the FLRW metric and two different metrics for $SU(1,1)$, $\bar{g}_1$ and $\bar{g}_2$.
\begin{eqnarray}
\label{metricsu11}
\bar{g}_1=dt \otimes dt-\frac{a(t)^2}{1-kr}dr \otimes dr-a(t)^2r^2\left(d\theta\otimes d\theta-\sin\theta^2 d \phi\otimes d \phi\right) \nonumber\\
+ 2b(t)^2(w^5 \otimes w^5-w^6 \otimes w^6-w^7\otimes w^7),
\end{eqnarray}
here,
\begin{equation}
2b(t)^2(w^5\otimes w^5-w^6\otimes w^6-w^7\otimes w^7)
\end{equation}
is a  bi-invariant metric for $SU(1,1)$ and
\begin{eqnarray}
\label{metrica2}
\label{metricsu112}
\bar{g}_2=dt \otimes dt-\frac{a(t)^2}{{1-kr}}dr \otimes  dr-a(t)^2r^2\left(d\theta\otimes d\theta-\sin\theta^2 d \phi\otimes d \phi \right)\nonumber\\+ 2b(t)^2(w^5 \otimes w^5)-2c(t)^2(w^6 \otimes w^6-w^7\otimes w^7),
\end{eqnarray}
where
\begin{equation}
2b(t)^2(w^5\otimes w^5)-2c(t)^2(w^6\otimes w^6+w^7\otimes w^7)
\end{equation}
 is a left invariant metric, $\left\{w^5,w^6,w^7\right\}$ is a basis for the left-invariant 1-forms of $SU(1,1)$, $a(t)$ is the scale factor and $b(t)$ is a function dependent of time. From (\ref{metricsu11}) and (\ref{metricsu112}) we can see that there are two temporal dimensions, this is because  a bi-invariant metric is constructed by means of the structure constants of the group. We  can also consider a metric with only  extra spatial dimensions, that is to say in (\ref{metricsu11}) and (\ref{metricsu112}), we can replace the  positive sign of the extra dimension by a negative one, and we can find the field equations, but for the first case the field equations are not consistent and for the second we have a decelerating universe.
 
 So, in this section we get the field equations for $\bar{g}_1$ and $\bar{g}_2$ and propose an equation of state for that what we call "dark energy". 

\subsection{Equations for $\bar{g}_1$ and $\bar{g_2}$}

In this section we are going to obtain the field  equations of $\bar{g}_2$. The field equations of $\bar{g_1}$ can be obtained when we replace $b(t)=c(t)$.
Again, we use the structural equations of Cartan to find the non-vanishing components of the Einstein's tensor. In this case,
\begin{eqnarray}
\theta^1 &\equiv & dt \nonumber\\
\theta^2 &\equiv &\frac{a(t)}{1-kr} dr\nonumber\\ 
\theta^3 &\equiv & a(t)r d\theta\nonumber\\
\theta^4 &\equiv &a(t)r \sin\theta d\phi\nonumber\\
\theta^5 &\equiv &\sqrt{2}b(t)w^5\nonumber\\
\theta^6 &\equiv &\sqrt{2}b(t)w^6\nonumber\\
\theta^7 &\equiv &\sqrt{2}c(t)w^7
\end{eqnarray}
forms the dual basis of an orthonormal basis, such that, $\bar{g}=\eta_{AB}\theta^A\otimes \theta^B$, with
\begin{equation}
(\eta_{AB})=\text{diag}\left\{1,-1,-1,-1,1,-1,-1\right\}
\end{equation}
To obtain the Einstein's equations we use  a diagonal energy momentum tensor of a perfect fluid with a barotropic equation of state, $p=w\rho$,
\begin{equation}
\label{tensorenergia7}
T_{AB}=\text{diag}(\rho(t),p(t),p(t),p(t),0,0,0),
\end{equation}
where $\rho(t)$ is the density and $p(t)$ is  the pressure, we arrive at
\begin{subequations}
\begin{eqnarray}
\label{E1SU11}
3\frac{\dot{a}^2}{a^2}+\frac{\dot{c}^2}{c^2}+3\frac{\dot{a}\dot{b}}{ab}+6\frac{\dot{a}\dot{c}}{ac}+2\frac{\dot{b}\dot{c}}{bc}+\frac{3k}{a^2}-\frac{1}{2c^2}+\frac{b^2}{8c^4}&=&\kappa\rho\\
\label{E2SU11}
2\frac{\ddot{a}}{a}+\frac{\ddot{b}}{b}+2\frac{\ddot{c}}{c}+\frac{\dot{a}^2}{a^2}+\frac{\dot{c}^2}{c^2}+2\frac{\dot{a}\dot{b}}{ab}+4\frac{\dot{a}\dot{c}}{ac}+2\frac{\dot{b}\dot{c}}{bc}+\frac{k}{a^2}-\frac{1}{2c^2}+\frac{b^2}{8c^4}&=&-\kappa p\\
\label{E3SU11}
3\frac{\ddot{a}}{a}+2\frac{\ddot{c}}{c}+3\frac{\dot{a}^2}{a^2}+\frac{\dot{c}^2}{c^2}+6\frac{\dot{a}\dot{c}}{ac}+3\frac{k}{a^2}-\frac{1}{2c^2}+\frac{3}{8}\frac{b^2}{c^4}&=&0\\
\label{E4SU11}
3\frac{\ddot{a}}{a}+\frac{\ddot{b}}{b}+\frac{\ddot{c}}{c}+3\frac{\dot{a}^2}{a^2}+3\frac{\dot{a}\dot{c}}{ac}+\frac{\dot{b}\dot{c}}{bc}+3\frac{\dot{a}\dot{b}}{ab}+3\frac{k}{a^2}-\frac{b^2}{8c^4}&=&0\\
\label{densi2}
\dot{\rho}+3\frac{\dot{a}}{a}(\rho+p)+\left(\frac{\dot{b}}{b}+2\frac{\dot{c}}{c}\right)\rho&=&0
\end{eqnarray}
\end{subequations}
We can integrate (\ref{densi2}) to obtain
\begin{equation}
\label{densityg2}
\rho(t)=\rho_0\frac{b_0c_0^2}{a^{3(1+w)}bc^2},
\end{equation}
where $\rho_0$, $b_0$ and $c_0$ are the present values of $\rho(t)$, $b(t)$ and $c(t)$ respectively.
 We can rewrite (\ref{E1SU11}) as
 \begin{equation}
 3\left(\frac{\dot{a}^2}{a^2}+\frac{k}{a^2}\right)=\kappa(\rho(t)+\rho_e),
 \end{equation}
 where
 \begin{equation}
 \kappa\rho_e=-3\frac{\dot{a}\dot{b}}{ab}-6\frac{\dot{a}\dot{c}}{ac}-2\frac{\dot{b}\dot{c}}{bc}+\frac{1}{2c^2}-\frac{b^2}{8c^4}
 \end{equation}
 and from equation (\ref{E2SU11}) we have
 \begin{eqnarray}
 2\left(\frac{\ddot{a}}{a}\right)+\left(\frac{\dot{a}}{a}\right)^2+\frac{k}{a^2}&=-\kappa p_{total}\nonumber\\
 &=-\kappa(p+p_e)\nonumber\\
 &=-\kappa(w\rho+\omega_e\rho_e),
 \end{eqnarray}
  where
 \begin{equation}
 \kappa\omega_e\rho_e=\frac{\ddot{b}}{b}+2\frac{\ddot{c}}{c}+\frac{\dot{c}^2}{c^2}+2\frac{\dot{a}\dot{b}}{ab}+4\frac{\dot{a}\dot{c}}{ac}+2\frac{\dot{b}\dot{c}}{bc}-\frac{1}{2c^2}+\frac{b^2}{8c^4},
 \end{equation}
 thus, if we consider that $p_e=\omega_e\rho_e$, then 
 \begin{equation}
 \label{omega1}
 \omega_e=\frac{\frac{\ddot{b}}{b}+2\frac{\ddot{c}}{c}+\frac{\dot{c}^2}{c^2}+2\frac{\dot{a}\dot{b}}{ab}+4\frac{\dot{a}\dot{c}}{ac}+2\frac{\dot{b}\dot{c}}{bc}-\frac{1}{2c^2}+\frac{b^2}{8c^4}}{-3\frac{\dot{a}\dot{b}}{ab}-6\frac{\dot{a}\dot{c}}{ac}-2\frac{\dot{b}\dot{c}}{bc}+\frac{1}{2c^2}-\frac{b^2}{8c^4}},
 \end{equation}
 but we can also rewrite $\omega_e$ as
 \begin{equation}
 \label{omega2}
     \omega_e=\frac{2\frac{\ddot{a}}{a}+\frac{\dot{a}^2}{a^2}+\frac{k}{a^2}+\kappa w\rho}{-3\frac{\dot{a}^2}{a^2}-3\frac{k}{a^2}+\kappa\rho}.
 \end{equation}
 This equation for $\omega_e$ in terms of $a$ and their derivatives is the same for any Lie group on the fiber, the difference is that $a$ is solution to their respective field equations.
 Here we have defined an equation of state to the "dark energy". Nevertheless, this dark energy is only a geometrical effect, that is to say, is a consequence of the topology of the universe.
 
 When $b(t)=c(t)$, we obtain the field equations for $\bar{g}_1$
\begin{subequations}
\begin{eqnarray}
\label{e1}
\frac{\dot{a}^2}{a^2}+\frac{\dot{b}^2}{b^2}+3\frac{\dot{a}\dot{b}}{ab}+\frac{k}{a^2}-\frac{1}{8b^2}&=&\kappa\frac{\rho(t)}{3}\\
\label{e2}
2\frac{\ddot{a}}{a}+3\frac{\ddot{b}}{b}+\frac{\dot{a}^2}{a^2}+\frac{6\dot{a}\dot{b}}{ab}+3\frac{\dot{b}^2}{b^2}+\frac{k}{a^2}-\frac{3}{8}\frac{1}{b^2}&=&-\kappa p(t)\\
\label{e3}
3\frac{\ddot{a}}{a}+2\frac{\ddot{b}}{b}+3\frac{\dot{a}^2}{a^2}+6\frac{\dot{a}\dot{b}}{ab}+\frac{\dot{b}^2}{b^2}+3\frac{k}{a^2}-\frac{1}{8b^2}&=&0
\end{eqnarray}
\end{subequations}
and according to (\ref{densityg2}) the density is
\begin{equation}
\label{densidadsu11}
\rho(t)=\frac{\rho_0b_0^3}{a^{3(1+w)}b^3}
\end{equation}
\begin{equation}
\omega_e=\frac{\kappa\rho_e}{\kappa p_e}=\frac{3\frac{\ddot{b}}{b}+6\frac{\dot{a}\dot{b}}{ab}+3\frac{\dot{b}^2}{b^2}-\frac{3}{8b^2}}{\frac{\dot{b}^2}{b^2}-9\frac{\dot{a}\dot{b}}{ab}-\frac{3}{8b^2}}=\frac{2\frac{\ddot{a}}{a}+\frac{\dot{a}^2}{a^2}+\frac{k}{a^2}+\kappa w\rho}{-3\frac{\dot{a}^2}{a^2}-3\frac{k}{a^2}+\kappa\rho}.
\end{equation}

A particular solution to the set of equations (\ref{e1}), (\ref{e2}), (\ref{e3}) is 
\begin{eqnarray}
\label{solution1}
a(t)=\left(\frac{t_0}{ t}\right)^{\frac{1}{3w+3}} ,\hspace{.5cm} b(t)=b_0\frac{t}{t_0},
\end{eqnarray}
with  $\frac{b_0^2}{t_0^2}=\frac{(3w+3)^2}{8(9w^2+9w+6)}$, being $t_0$ the present time. But if we replace this solution in (\ref{e1}), (\ref{e2}), (\ref{e3}), one finds that 
\begin{equation}
    \kappa\rho_0t_0^2=-\frac{15}{(3w+3)^2},
    \end{equation}
that is, the solution gives a negative density. Therefore we have to find different solutions for other initial conditions and since this system is highly coupled, we solved in section \ref{sec:solutions} the equations  (\ref{e1}), (\ref{e2}), (\ref{e3}) numerically as dynamical system. 

\section{Fiber bundle on the SU(2) group}
\label{section5}

Here we consider two different metrics for the group $SU(2)$, $\bar{g}_3$ and $\bar{g}_4$, the first  one is bi-invariant and the second one is left-invariant. 
 \begin{eqnarray}
 \bar{g}_3=dt \otimes dt-a(t)^2\left[\frac{dr \otimes  dr}{1-kr^2}+r^2(d\theta\otimes d\theta+\sin\theta^2 d \phi\otimes d \phi)\right]\nonumber\\- 2b(t)^2(w^5 \otimes w^5+w^6 \otimes w^6+ w^7\otimes w^7)
 \end{eqnarray}
 \begin{eqnarray}
 \bar{g}_4=dt \otimes dt-a(t)^2\left[\frac{dr \otimes  dr}{1-kr^2}+r^2(d\theta\otimes d\theta+\sin\theta^2 d \phi\otimes d \phi)\right]\nonumber\\-2b(t)^2(w^5 \otimes w^5)-2c(t)^2(w^6 \otimes w^6+w^7\otimes w^7)
 \end{eqnarray}
 The Einstein's equations using $\bar{g}_3$ are
\begin{subequations}
\begin{eqnarray}
\label{esu21}
\frac{\dot{a}^2}{a^2}+\frac{\dot{b}^2}{b^2}+3\frac{\dot{a}\dot{b}}{ab}+\frac{k}{a^2}+\frac{1}{8b^2}&=&\kappa\frac{\rho(t)}{3}\\
\label{esu22}
2\frac{\ddot{a}}{a}+3\frac{\ddot{b}}{b}+\frac{\dot{a}^2}{a^2}+\frac{6\dot{a}\dot{b}}{ab}+3\frac{\dot{b}^2}{b^2}+\frac{k}{a^2}+\frac{3}{8}\frac{1}{b^2}&=&-\kappa p(t)\\
\label{esu23}
3\frac{\ddot{a}}{a}+2\frac{\ddot{b}}{b}+3\frac{\dot{a}^2}{a^2}+6\frac{\dot{a}\dot{b}}{ab}+\frac{\dot{b}^2}{b^2}+3\frac{k}{a^2}+\frac{1}{8b^2}&=&0.
\end{eqnarray}
\end{subequations}
The density is given by (\ref{densidadsu11}). Einstein's equations using $\bar{g}_4$ are
 \begin{subequations}
 \begin{eqnarray}
3\frac{\dot{a}^2}{a^2}+\frac{\dot{c}^2}{c^2}+3\frac{\dot{a}\dot{b}}{ab}+6\frac{\dot{a}\dot{c}}{ac}+2\frac{\dot{b}\dot{c}}{bc}+\frac{3k}{a^2}-\frac{1}{8}\frac{b^2}{c^4}+\frac{1}{2c^2}&=&\kappa\rho\\
2\frac{\ddot{a}}{a}+\frac{\ddot{b}}{b}+2\frac{\ddot{c}}{c}+\frac{\dot{a}^2}{a^2}+\frac{\dot{c}^2}{c^2}+2\frac{\dot{a}\dot{b}}{ab}+4\frac{\dot{a}\dot{c}}{ac}+2\frac{\dot{b}\dot{c}}{bc}+\frac{k}{a^2}-\frac{1}{8}\frac{b^2}{c^4}+\frac{1}{2c^2}&=&-\kappa p\\
3\frac{\ddot{a}}{a}+2\frac{\ddot{c}}{c}+3\frac{\dot{a}^2}{a^2}+\frac{\dot{c}^2}{c^2}+6\frac{\dot{a}\dot{c}}{ac}+\frac{3k}{a^2}-\frac{3}{8}\frac{b^2}{c^4}+\frac{1}{2c^2}&=&0\\
3\frac{\ddot{a}}{a}+\frac{\ddot{b}}{b}+\frac{\ddot{c}}{c}+3\frac{\dot{a}^2}{a^2}+3\frac{\dot{a}\dot{c}}{ac}+3\frac{\dot{a}\dot{b}}{ab}+\frac{\dot{b}\dot{c}}{bc}+\frac{3k}{a^2}+\frac{b^2}{8c^4}&=&0.
\end{eqnarray}
\end{subequations}
The density is again given by (\ref{densityg2}) and we can also define an equation of state for the "dark energy" like in the previous section. 

\section{Solutions}
\label{sec:solutions}

In this section, we solve the field equations for $\bar{g}_1$, $\bar{g}_2$, $\bar{g}_3$ and $\bar{g}_4$, numerically, considering $k=0$. For doing that, we rewrite them in terms of $X=H_0t$ and of the density parameter.  
Observe that if the present value of $a_0=1=a(t_0)$, where $t_0$ is the present time, then $a_0'=\frac{1}{H_0}\dot{a}_0=1$, where we have considered that $\dot{a}_0=H_0$. In general for the solutions that we find $a=1$ for a $t\neq t_0$. Because we want solutions that resemble our Universe, we look for solutions  such that when $a=1$ then  $a'=1$, $b=b_0$ and $c=c_0$. In this way when $a=1$ the velocity of expansion  is equal to $H_0$ and the density is $\rho_0$.
For each metric we find a set of equations that we can solve as a dynamical system for different initial conditions,  using an appropriated semi-implicit extrapolations method for the resulting stiff system. We take the initial value of $a$ as $0.001$, that means that our solutions are valid from a redshift $z=999$. 
\subsection{Solutions for the metric $\bar{g}_1$}

In terms of the new parameters, from (\ref{e1}) to (\ref{e3}) we get the following expressions:
\begin{eqnarray}
\label{asu11}
a''&=&\frac{\Omega_{0\gamma}b_0^3}{3a^3b^3}+\frac{\Omega_{0m}b_0^3}{5a^2b^3}+\frac{5}{2}\frac{a'^2}{a} \mp\frac{3}{2}\frac{a'}{b}\left(5\frac{a'^2}{a^2}b^2+\frac{1}{2H_0^2}+\frac{4}{3}\left(\frac{\Omega_{0m}b_0^3}{a^3b}+\frac{\Omega_{0\gamma}b_0^3}{a^4b}\right)\right)^\frac{1}{2}\nonumber\\
\label{bsu11}
b'&=&-\frac{3}{2}\frac{a'}{a}b\pm\frac{1}{2}\left(5\frac{a'^2}{a^2}b^2+\frac{1}{2H_0^2}+\frac{4}{3}\left(\frac{\Omega_{0m}b_0^3}{a^3b^3}+\frac{\Omega_{0\gamma}}{a^4b^3}\right)\right)^\frac{1}{2}.
\end{eqnarray}
These two equations are solved as a dynamical system.  To do that, we choose a point in the past, $i.e.$, we choose initial conditions denoting the quantities related to this point with the index $i$. We take $a_i=0.001$ and $a_i'>0$, $b_i\neq 0$, $b_0\neq0$ and, according to (\ref{densidadsu11}), we take $b_i$ and $b_0$ with  the same sign.
From (\ref{asu11}), considering the negative sign for $a''$, we can see that this equation permits that there can be initial accelerating or decelerating expansion, it should depend on the initial conditions. Nevertheless, numerically we find that there are only solutions for an initial decelerating expansion and that $a''$ remains negative throughout all the time, that is, there is always a decelerating expansion.
Meanwhile, if we consider the positive sign in  (\ref{asu11}), we can see that no matter what initial conditions we choose, we can always get an initial accelerating expansion, but this case doesn't have numerical solutions. So this metric only gives rise to a decelerating expansion. 

\subsection{Solutions for the metric $\bar{g}_2$}

\begin{figure}[htbp]
\subfigure[$a$.  ]
{\includegraphics[width=9cm]{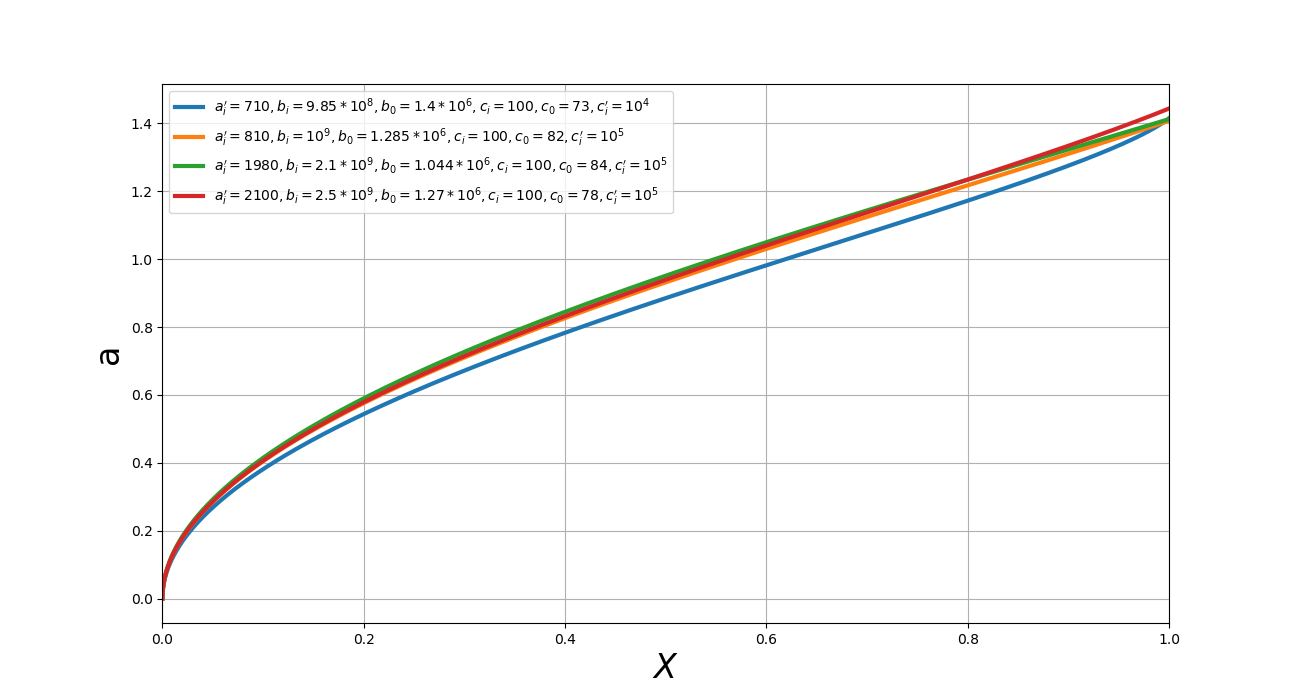}}
\subfigure[ $b$.]
{\includegraphics[width=9cm]{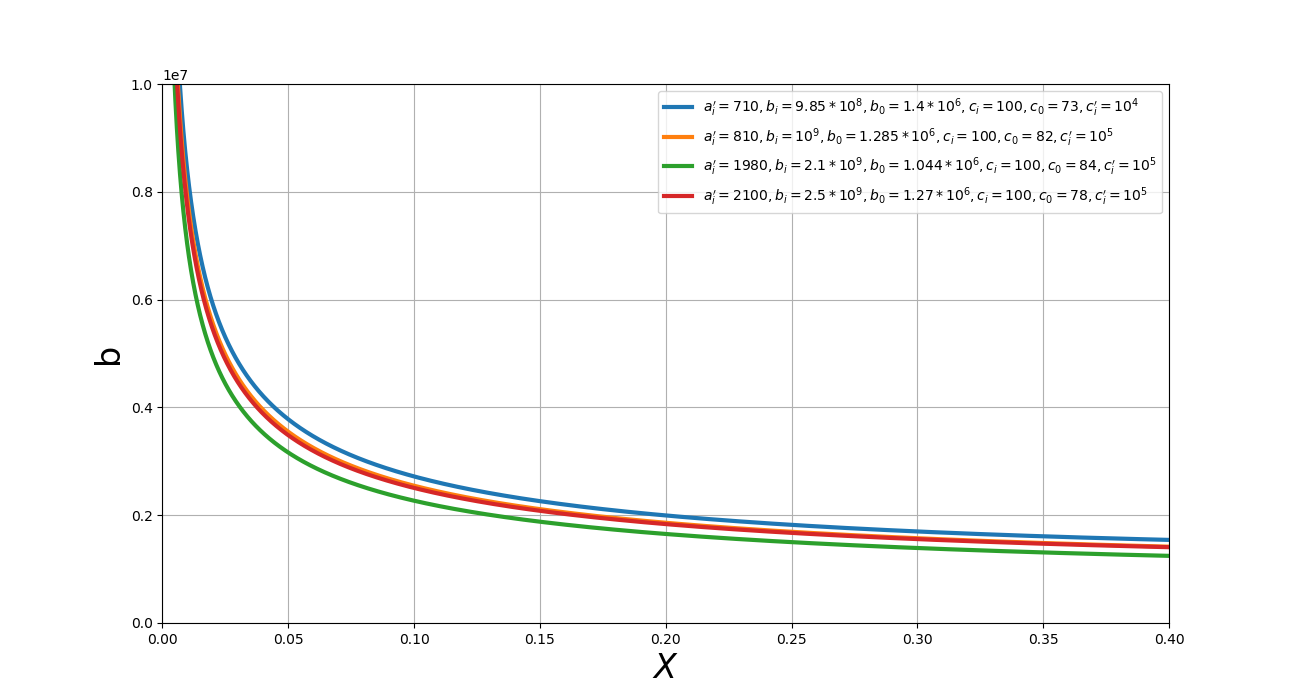}}
\subfigure[$a''$] {\includegraphics[width=9cm]{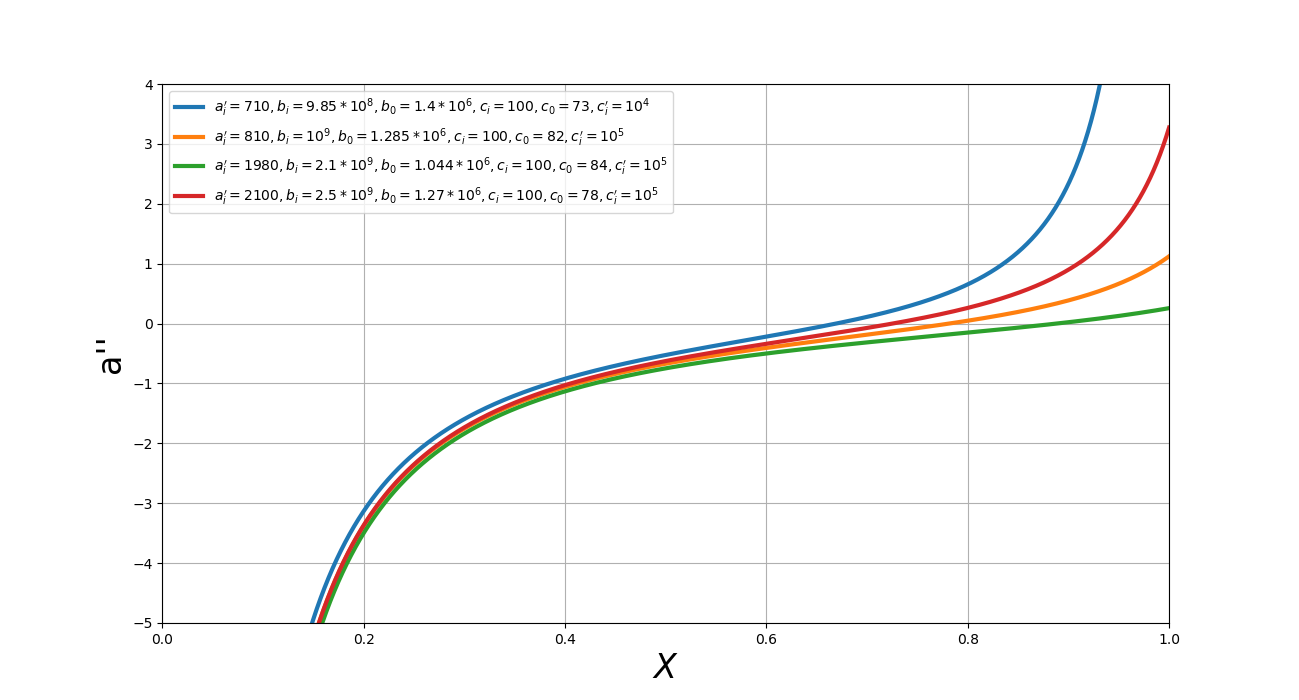}}
\subfigure[$c$]
{\includegraphics[width=9cm]{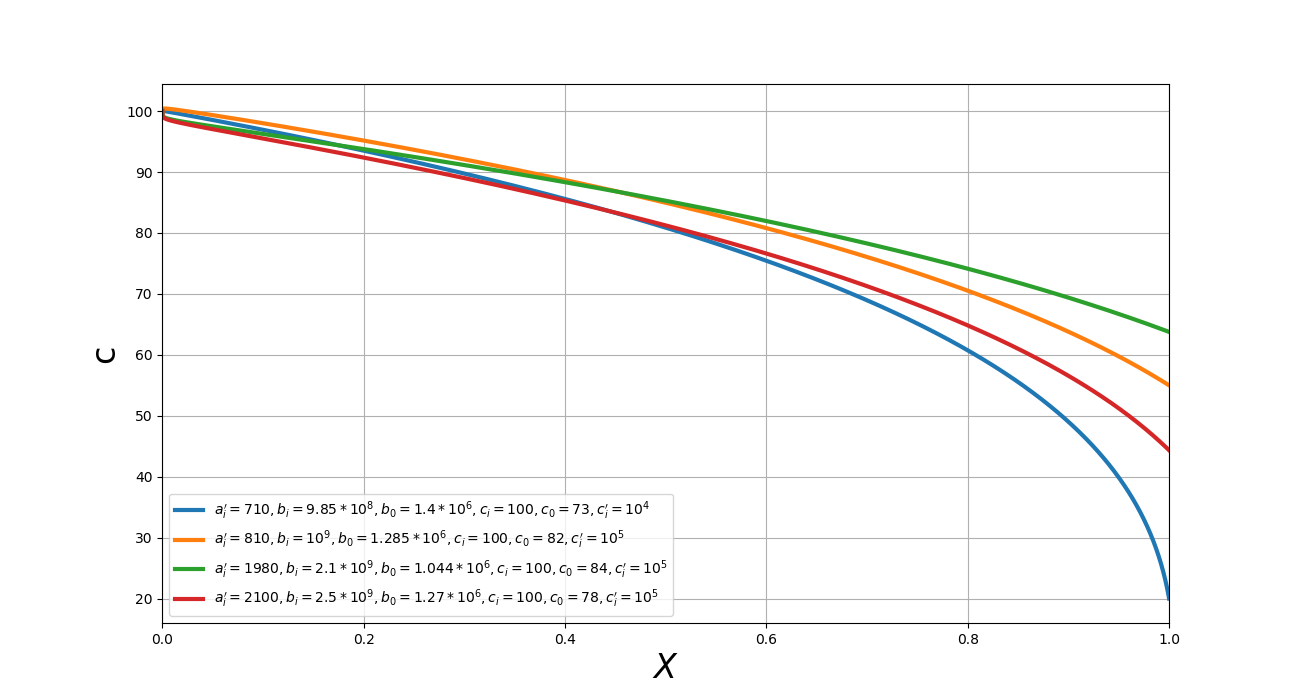}}
\subfigure[ $\omega_e$ ]
{\includegraphics[width=9cm]{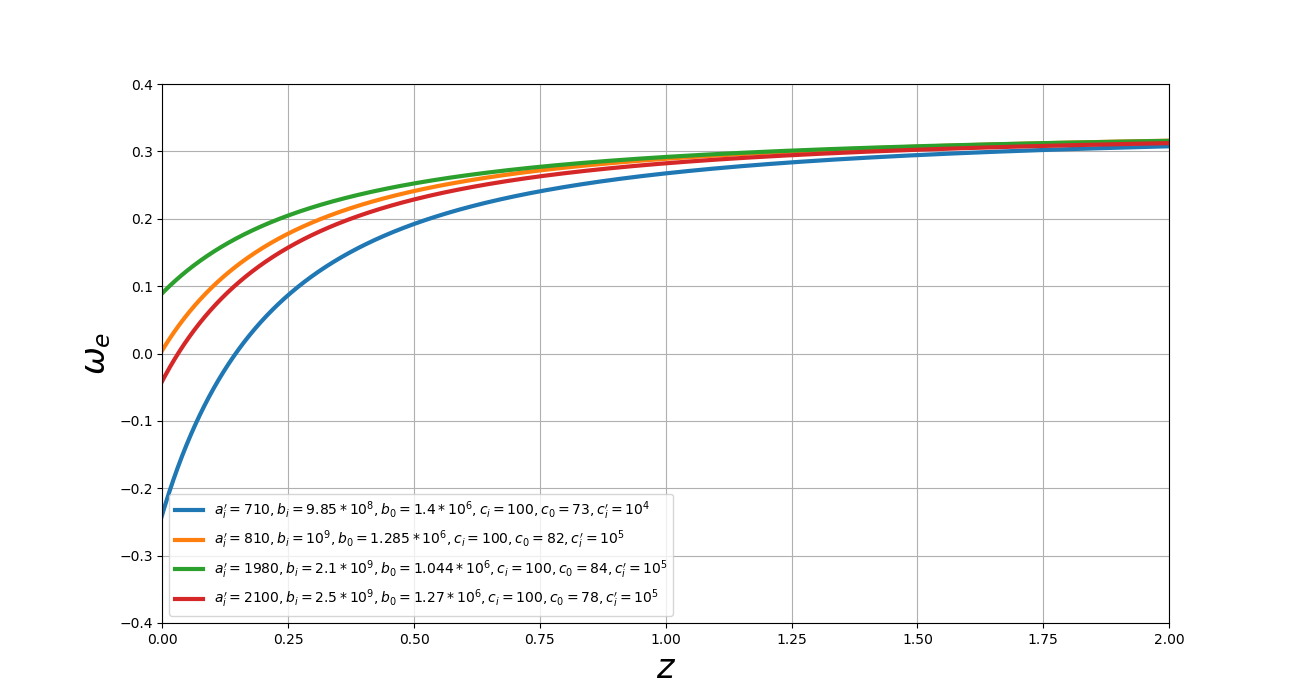}}
\caption{Evolution of  the scale factors  $a$, $b$, $c$ and the EoS $\omega_e$ when the fiber is the group $SU(1,1)$, using metric $\bar{g}_2$, for different initial conditions that give rise to an initial decelerating expansion that, at some point, becomes accelerating. 
(a)  Evolution of the scale factor $a$. From top to bottom, $a=1$ in $X=0.57$, $0.54$, $0.57$ and $0.61$. So with $SU(1,1)$ as fiber, 
the scale factor grows faster than the observed one. 
(b) Evolution of the scale factor $b$. Observe that while the scale factor $a$ grows, the scale factor $b$ decays. 
(c) Acceleration of a. From top to bottom, the expansion becomes accelerated in $X=0.65,0.71,0.78$ and $0.88$. 
(d) Evolution of the scale factor $c$. As in the case of $b$, this scale factor also decays. 
(e) Evolution of the EoS $\omega_e$ for a redshift from $z=0$ to $z=2$.  The values of $\omega_e$ do not correspond to the observed ones. } \label{figura1}
\end{figure}

From (\ref{E1SU11}) to (\ref{E3SU11}) together with (\ref{densi2}) we find
\begin{eqnarray}
\label{asu1}
a''&=&\frac{2}{15}\frac{\Omega_{0\gamma}b_0c_0^2}{a^3bc^2}-\frac{7}{5}\frac{a'^2}{a}+\frac{1}{5}\frac{c'^2a}{c^2}-\frac{4}{5}\frac{a'c'}{c}+\frac{a}{H_0^2} \left(\frac{-1}{10c^2}+\frac{b^2}{40c^4}\right)+\\ &&\left[\frac{-3a'^2}{5a}\frac{-ac'^2}{5c^2}\frac{-6a'c'}{5c}+\frac{a}{10H_0^2c^2}-\frac{ab^2}{40H_0^2c^4}+\frac{\Omega_{0m}b_0 c_0^2}{5a^2bc^2}+\frac{\Omega_{0\gamma}b_0c_0^2}{5a^3bc^2}\right]\left(\frac{2c'a-2a'c}{3a'c+2c'a}\right)\nonumber\\ \nonumber\\
 \label{bsu1}
b'&=&\frac{\left[\frac{-3a'^2bc}{a}-\frac{ac'^2b}{c}-6a'c'b+\frac{1}{H_0^2}\left(\frac{ab}{2c}-\frac{ab^3}{8c^3}\right)+\frac{\Omega_{0m}b_0c_0^2}{a^2c}+\frac{\Omega_{0\gamma}b_0c_0^2}{a^3c}\right]}{3a'c+2c'a}\\ \nonumber\\
\label{csu1}
c''&=&\frac{3}{5}\frac{a'^2c}{a^2}-\frac{4c'^2}{5c}-\frac{9a'c'}{5a}+\frac{1}{H_0^2}\left(\frac{2}{5c}-\frac{9b^2}{40c^3}\right)-\frac{\Omega_{0\gamma}b_0c_0^2}{5a^4bc}+\\
&&\left[\frac{-3a'^2c}{5a^2}-\frac{c'^2}{5c}-\frac{6a'c'}{5a}+\frac{1}{10H_0^2c}-\frac{b^2}{40H_0^2c^3}+\frac{\Omega_{0m}b_0c_0^2}{5a^3bc}+\frac{\Omega_{0\gamma}b_0c_0^2}{5a^4bc}\right]\left(\frac{3a'c-3c'a}{3a'c+2c'a}\right)\nonumber
\end{eqnarray}
 We solved (\ref{asu1}), (\ref{bsu1}), (\ref{csu1}) again as a dynamical system for different initial conditions.
We  find that there can be universes with an initial decelerating expansion and that its magnitude decreases with time and tends to zero, but  there are also initial conditions that give rise to an initial decelerating expansion, that, in some point of time, becomes accelerating, that is to say, there is a point where the universe starts to accelerate. This point where it starts to accelerate depends on the initial conditions, as we can see in Figure \ref{figura1}c. In all the figures, the value of $X=1$ represents approximately the age of the universe. In Figure  \ref{figura1}b we can see that the scale factor is equal to 1 for $X<1$. So with this metric, the scale factor grows up faster than the observed.  In Figure \ref{figura1}e we  plot the evolution of $\omega_e$ for a redshift from $z=0$ to $z=2$. As we can see the value of $\omega_e$  depends on the initial velocity of  expansion of the scale factor (\ref{omega2}). $\omega_e$  can take negative values but it is far from the present value of $\omega_e$ $<-1$. As we can see, when $z=2$ the value of $\omega_e$ is approximately the same, $\omega_e\approx .3$, for different initial conditions. This is also true for a redshift $z>2$. 

On the other hand,  (\ref{omega2})  can be written in terms of $X$ as
 \begin{equation}
 \label{omega3}
     \omega_e=\frac{\frac{2a''}{a}+\frac{a'^ 2}{a^2}+\frac{k}{a^2}+\frac{\Omega_{0\gamma}b_0c_0^2}{3a^4bc^2}}{-3\frac{a'^2}{a^2}-3\frac{k}{a^2}+\frac{\Omega_{0m}b_0c_0^2}{a^3bc^2}+\frac{\Omega_{0\gamma}}{a^4b^2}}.
 \end{equation}

As we mentioned before, we look for solutions where $a=1$ implies $a'=1$, $b=b_0$ and $c=c_0$. Then from (\ref{omega3}), when $a=1$ we have
\begin{equation}
\label{omega6}
    \omega_e=\frac{-2a''-1-\frac{\Omega_{0\gamma}}{3}}{3-\Omega_{0m}-\Omega_{0\gamma}}.
\end{equation}
If we replace the present values of $\Omega_{0\gamma}$ and $\Omega_{0m}$ in equation (\ref{omega6}), we can see that
 if $a''>0$ when $a=1$, then $\omega_e$ is negative, but if $a''<0$ when $a=1$, then $\omega_e$ can be positive or zero. As we know $a(t)=\frac{1}{1+z}$, so $a=1$ is equivalent to $z=0$. Therefore, if $a''>0$ when $a=1$, then $\omega_e<0$ in $z=0$, but if $a''<0$ when $a=1$, then $\omega_e\geq 0$ in $z=0$, being this true for any Lie group. So, this tell us that if our universe is topologically a principal fiber bundle and there is a current accelerating expansion, then the current value of $\omega_e$ has to be negative. 
\begin{figure}[htbp]
\subfigure[$a$]
{\includegraphics[width=9cm]{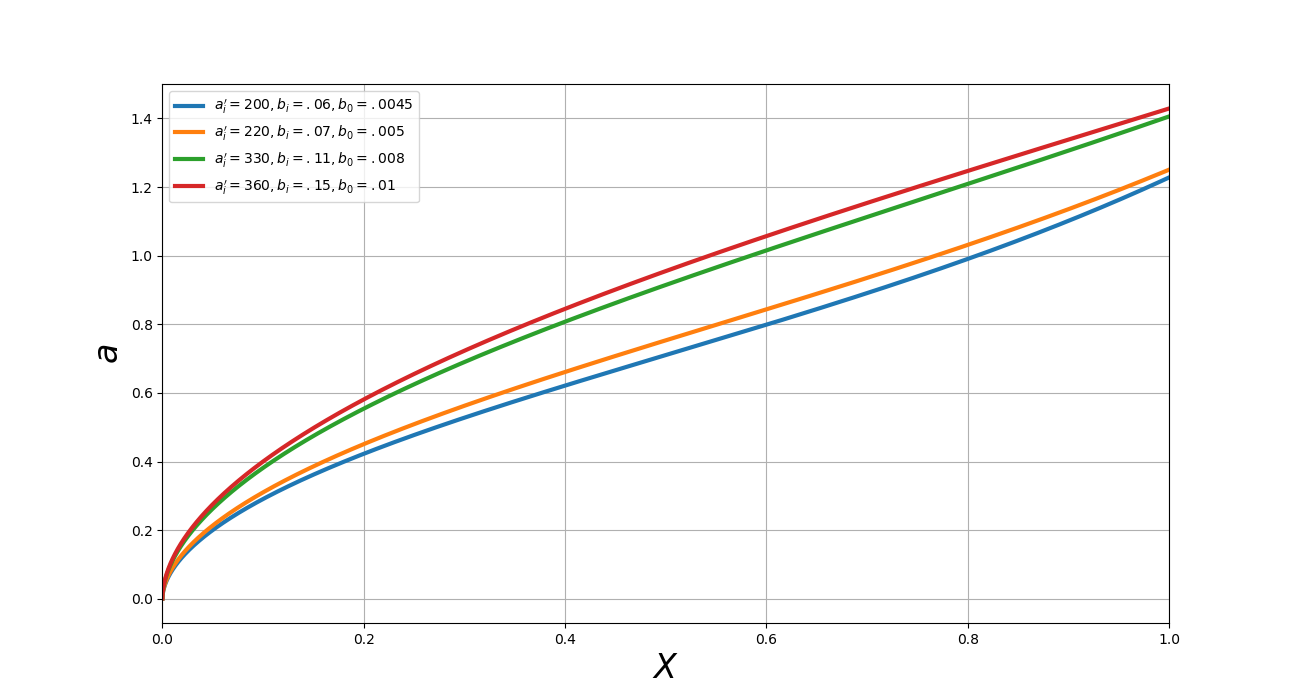}}
\subfigure[$a''$]
{\includegraphics[width=9cm]{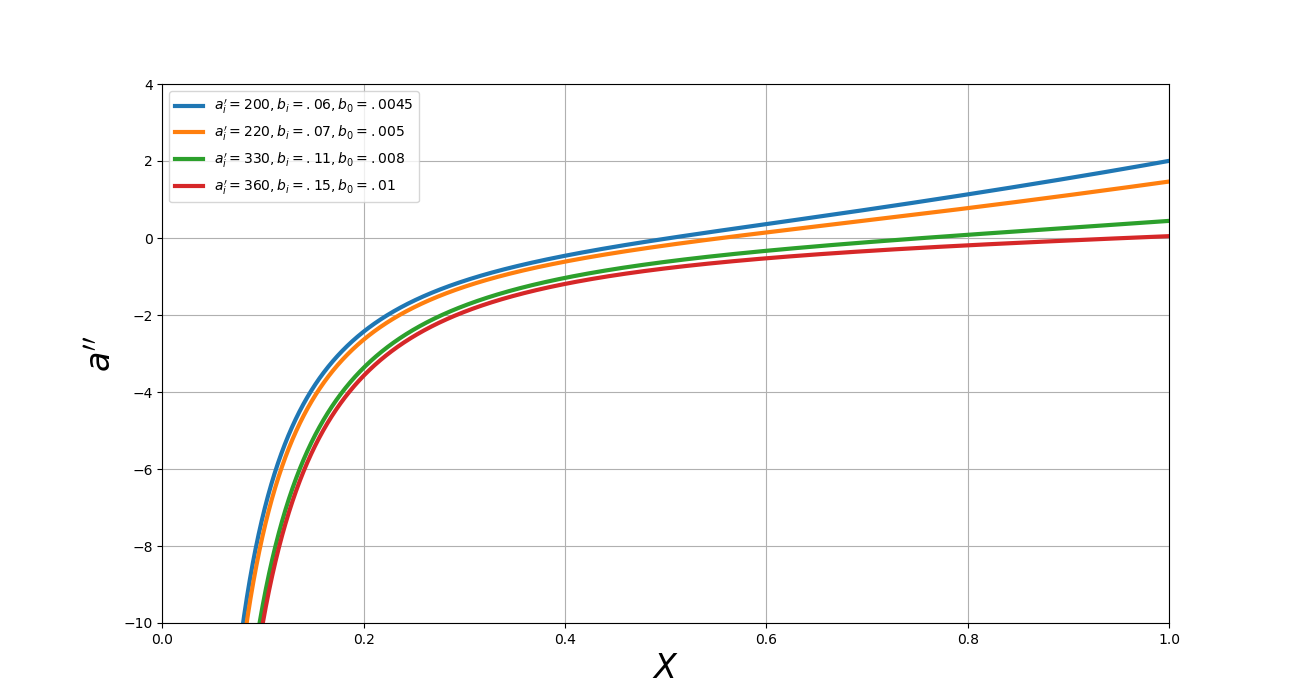}}
\subfigure[$b$]
{\includegraphics[width=9cm]{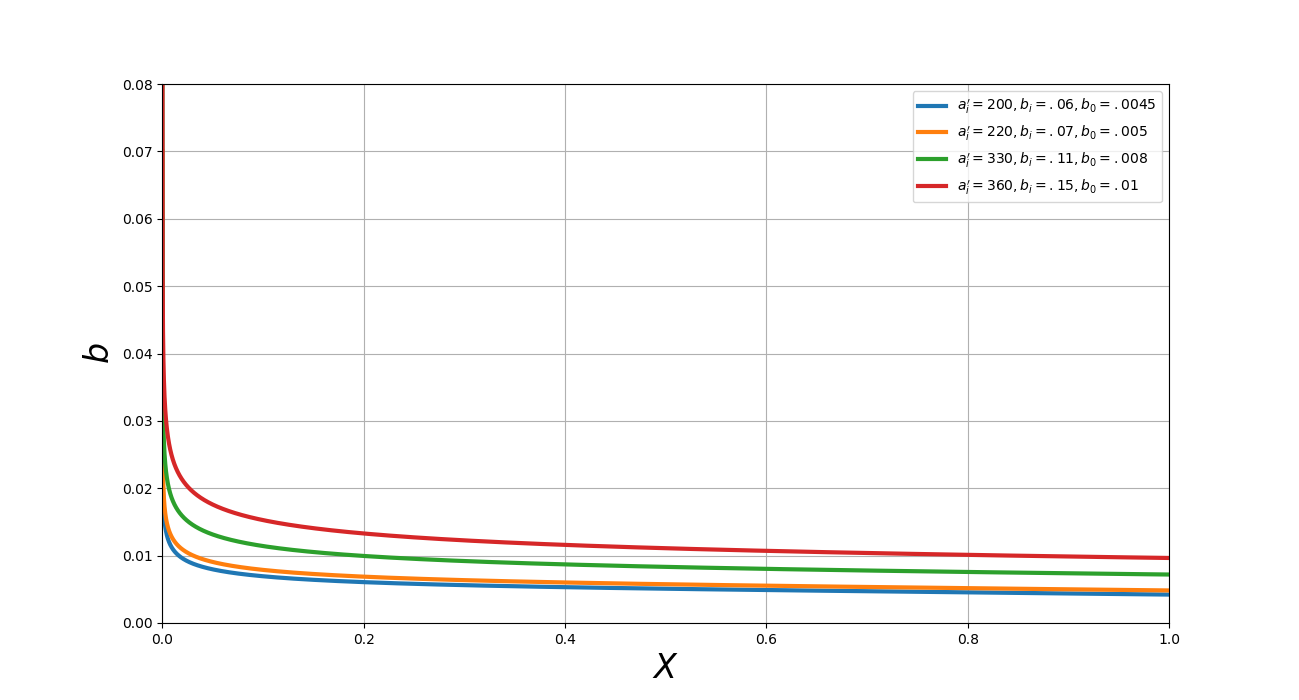}}
\subfigure[$\omega_e$]{\includegraphics[width=9cm]{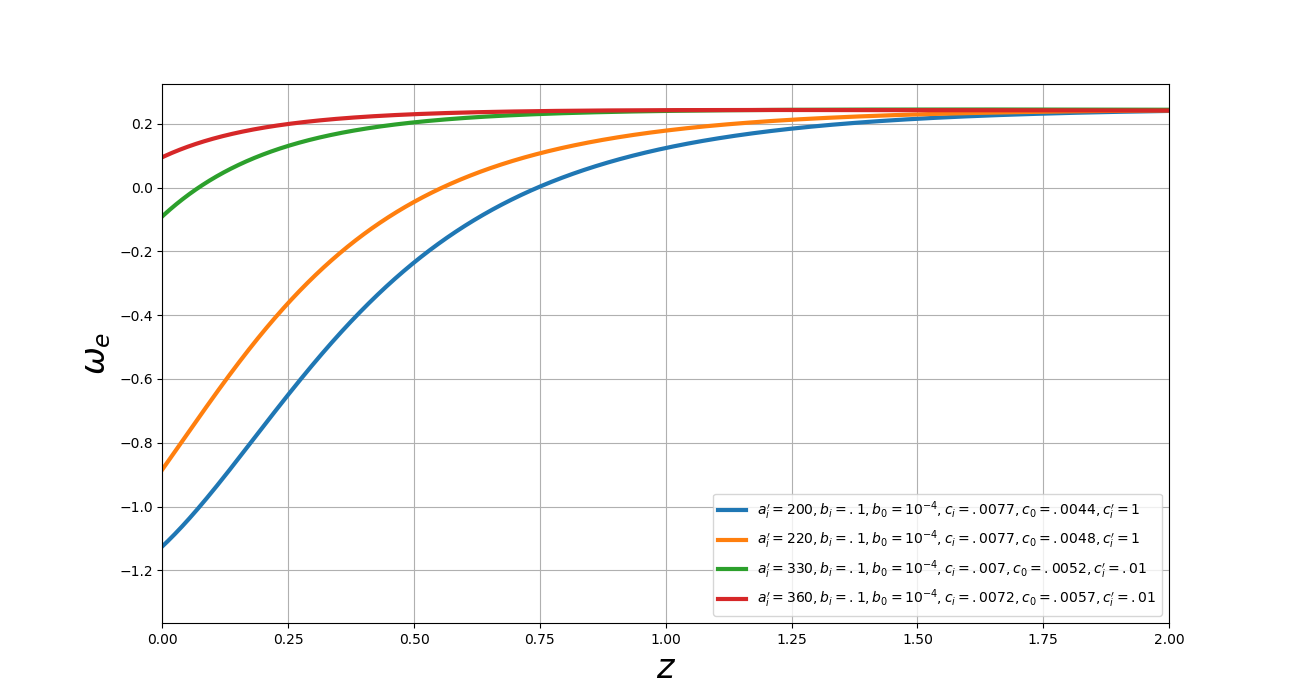}}
\caption{Evolution of the scale factors $a$, $b$, $c$ and the EoS $\omega_e$ when the fiber is the group $SU(2)$, using metric $\bar{g}_3$, for different initial conditions that give rise to an initial decelerating expansion that, at some point, becomes accelerating. 
(a) Evolution of $a$. From top to bottom: $a=1$ in $X=0.53, 0.58, 0.76$ and $0.8$. 
(b) Evolution of the acceleration of $a$. From top to bottom, the expansion becomes accelerated in $X=0.5, 0.55, 0.75$ and $0.95$. 
(c) Evolution of the scale factor $b$. Observe again that while the scale factor $a$ grows, the scale factor $b$ decays.
(d) The evolution of the EoS $\omega_e$.
As we can see, there are initial conditions  that allow $\omega_e$ to have values $<-1$.}
\label{figura2}
\end{figure}

\subsection{Solutions for the metric $\bar{g}_3$}

In the same way, for the metric $\bar{g}_3$ we have
\begin{eqnarray}
\label{acesu2}
a''&=&\frac{\Omega_{0\gamma}b_0^3}{3a^3b^3}+\frac{\Omega_{0m}b_0^3}{5a^2b^3}+\frac{5}{2}\frac{a'^2}{a}
\mp\frac{3}{2}\frac{a'}{b}\left(5\frac{a'^2}{a^2}b^2-\frac{1}{2H_0^2}+\frac{4}{3}\left(\frac{\Omega_{0m}b_0^3}{a^3b}+\frac{\Omega_{0\gamma}b_0^3}{a^4b}\right)\right)^\frac{1}{2}
\\
\label{b'su2}
b'&=&-\frac{3}{2}\frac{a'}{a}b\pm\frac{1}{2}\left(5\frac{a'^2}{a^2}b^2-\frac{1}{2H_0^2}+\frac{4}{3}\left(\frac{\Omega_{0m}b_0^3}{a^3b^3}+\frac{\Omega_{0\gamma}}{a^4b^3}\right)\right)^\frac{1}{2}
\end{eqnarray}
From (\ref{acesu2}), we can obtain the current acceleration given by: 
\begin{equation}
a_0''=\frac{\Omega_{0\gamma}}{3}+\frac{\Omega_{0m}}{5}+\frac{5}{2}\mp\frac{3}{2}\left(5-\frac{1}{2H_0^2b_0^2}+\frac{4}{3}\left(\Omega_{0m}+\Omega_{0\gamma}\right)\right)^\frac{1}{2}
\end{equation}
If we set the present value of $\Omega_{0m}\sim 0.28$ and $\Omega_{0\gamma}\sim 10^{-5}$, we see that in order to have real numbers for the square root, we have to set $b_0>0.0042367$.
So, we have a constraint for the value of $b_0$. Again, there are  two cases to be considered. The first one is when in (\ref{acesu2})  $a''$ has the positive sign and in (\ref{b'su2}) $b'$ is negative.  We find solutions for $a_i=0.001$, $b_0>.0042367$, $a_i'>0$ and, according to (\ref{densidadsu11}), $b_i>0$. We find that depending on the value of $b_0$ there are two kinds of solutions:

a) Decelerating expansion of the universe.

b) An initial decelerating expansion that, in some point, becomes accelerating. 

For the b case, we can see in Figure \ref{figura2}a that in general the scale factor grows faster than the observed one, but there are cases that grows less quickly compared to metric $\bar{g}_2$. In Figure \ref{figura2}b we plot $a''$.  We find  that $\omega_e$ can take negative values and that there are initial conditions where  $\omega_e<-1$ at $z=0$, as we can see in Figure \ref{figura2}d. This result is very interesting, because this dependence of $\omega_e$ with respect to $z$ is very similar to that found by recent observations (see \cite{Veto}). Also  we find that when $z\geq 2$ the value of $\omega_e$ is approximately of $.3$ for  different initial conditions. In Figure \ref{figura2}c we can see that, while the scale factor $a$ grows up,  $b$ decreases.  

Now we study the second case, when we consider in (\ref{acesu2}) the positive sign and in (\ref{b'su2}) the negative sign. We can calculate $a_i''$ for some point when $a_i=0.001$ and $a_i'>0$ but this give a positive initial acceleration for any value of $b_0$ and $b_i$. So, this solution is not physical and furthermore doesn't have numerical solutions.

\subsection{Solutions for the metric $\bar{g}_4$}

\begin{figure}[htbp]
\subfigure[$a$]{\includegraphics[width=9cm]{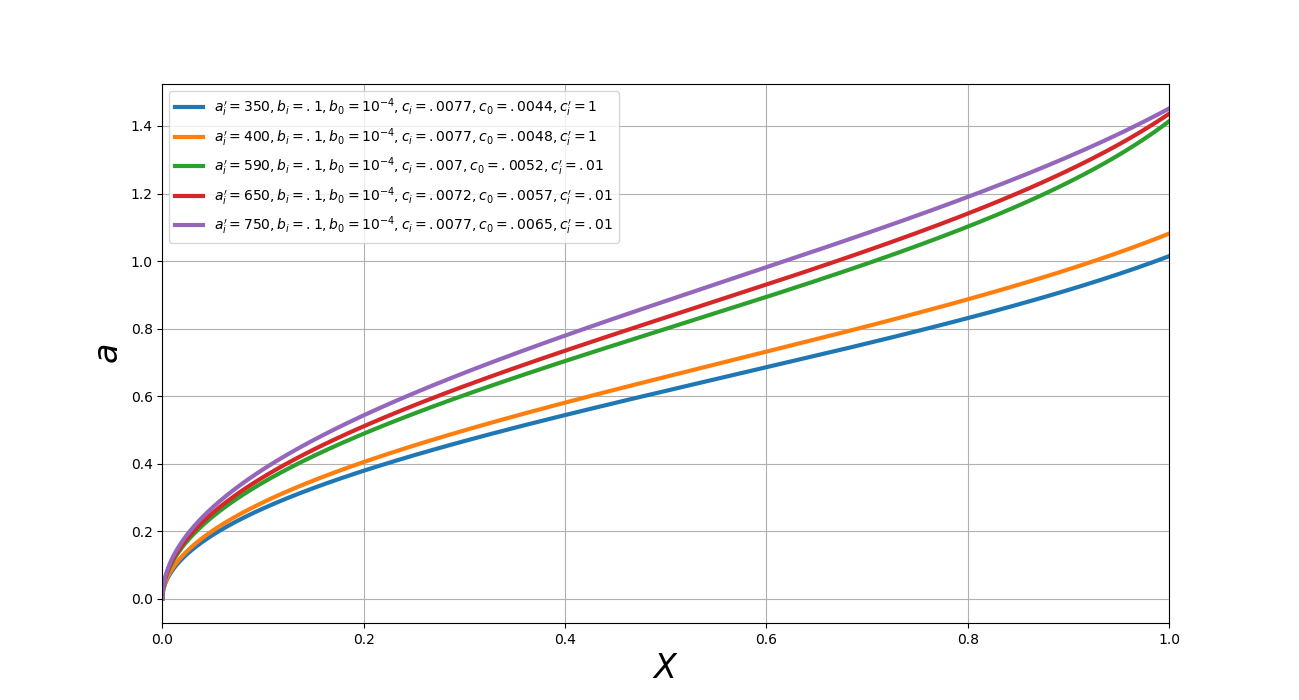}}
\subfigure[$a''$]{\includegraphics[width=9cm]{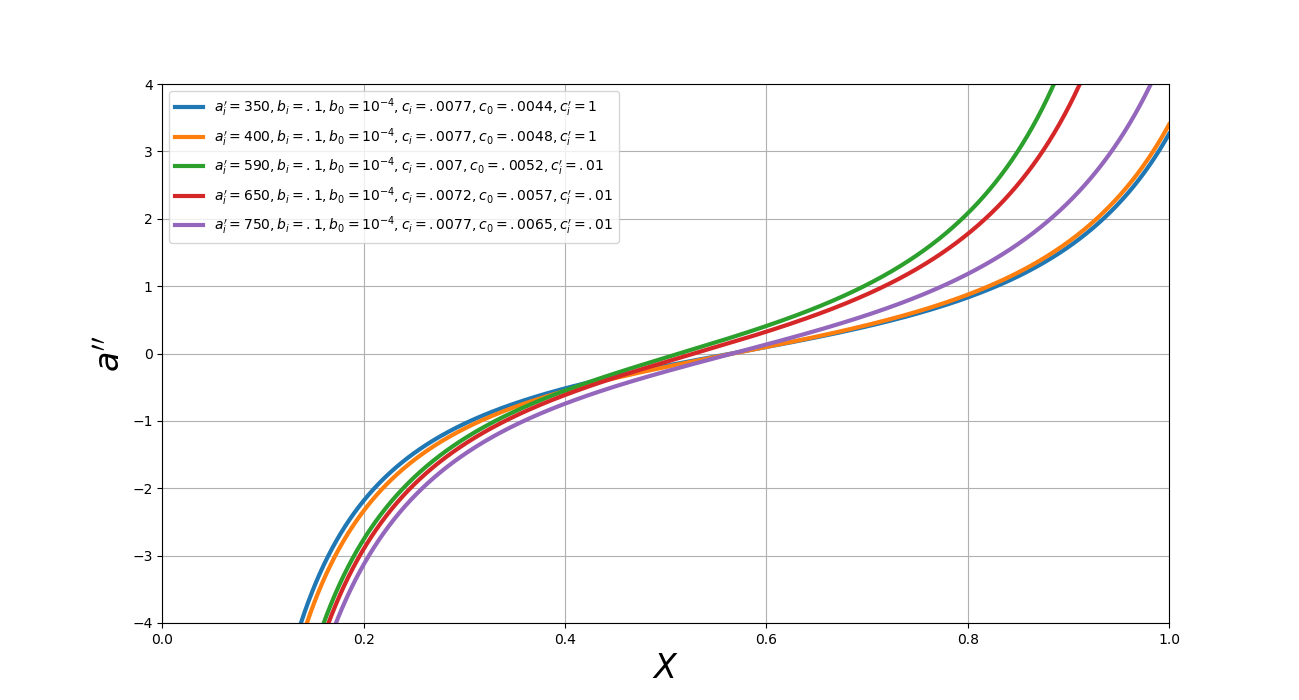}}
\subfigure[$b$]{\includegraphics[width=9cm]{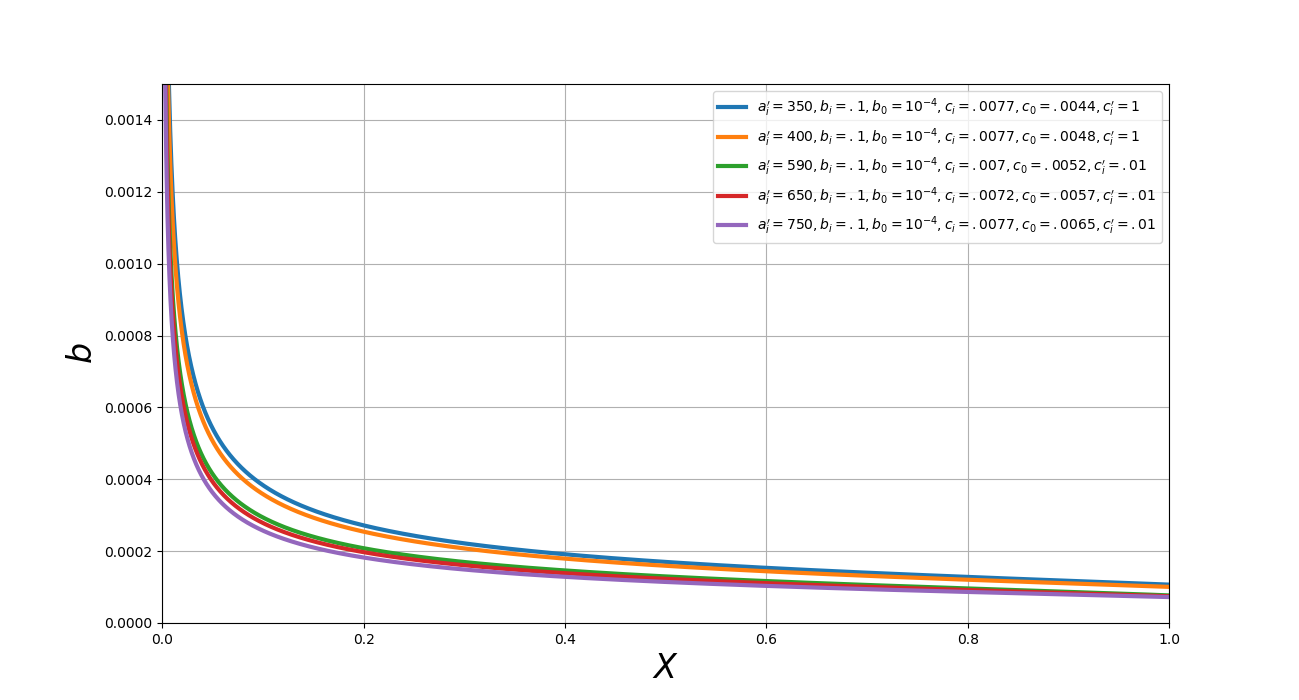}}
\subfigure[$c$]{\includegraphics[width=9cm]{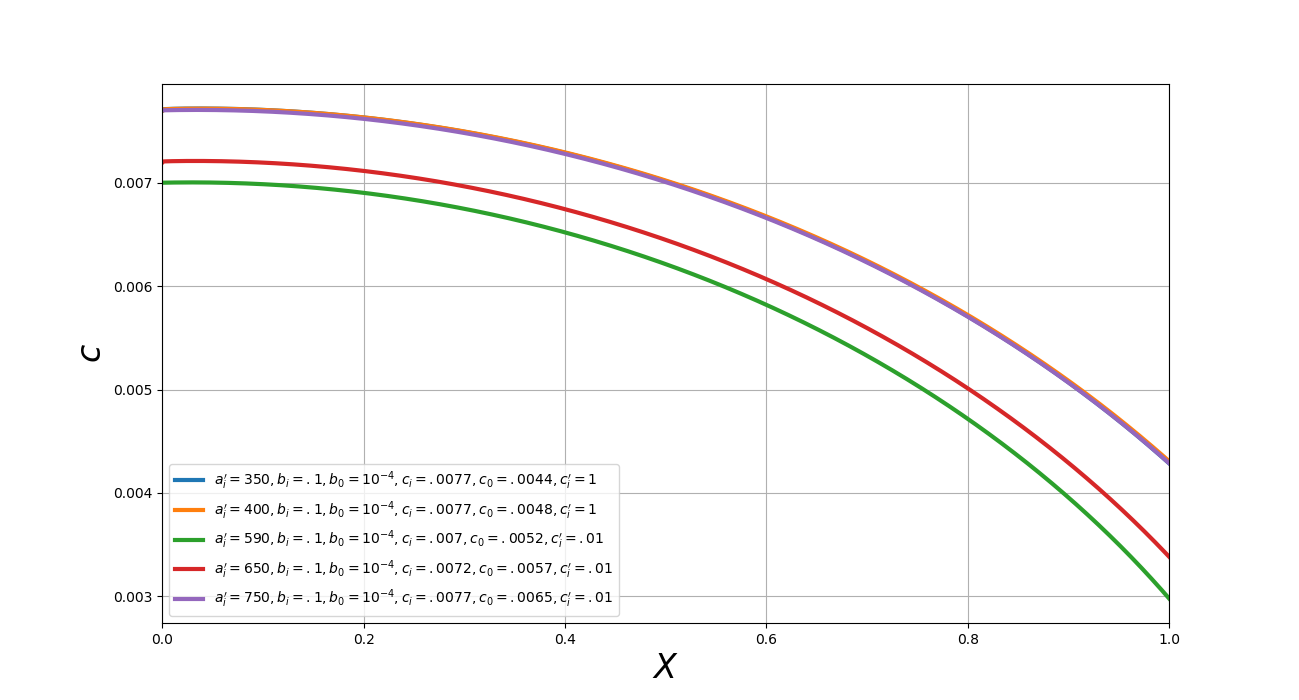}}
\subfigure[$\omega_e$]{\includegraphics[width=9cm]{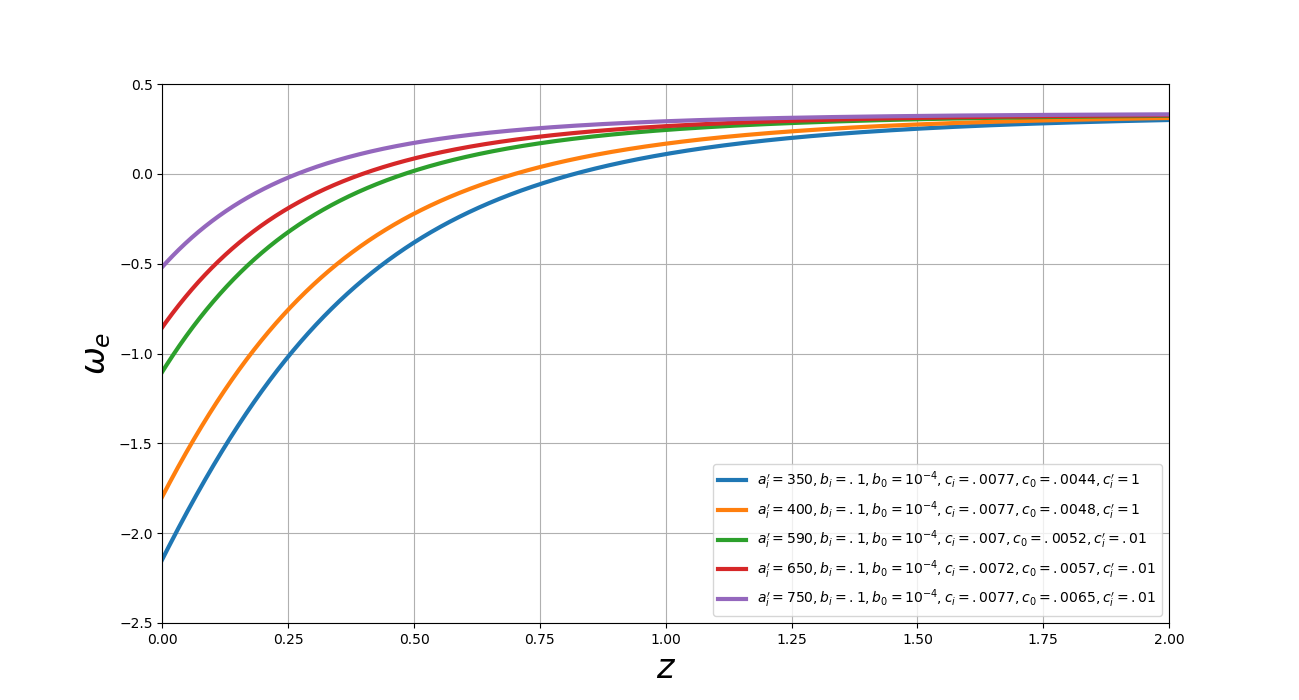}}
\caption{Evolution of the scale factor $a$, $b$, $c$ and the EoS $\omega_e$ when the fiber is the group $SU(2)$, using metric $\bar{g}_4$, for different initial conditions that give rise to an initial decelerating expansion that, at some point, becomes accelerating. 
(a) Evolution of the scale factor $a$. From top to bottom $a=1$ in $X=0.61, 0.66, 0.7, 0.92$ and $0.98$.
(b) Acceleration of $a$. From top to bottom, the scale factor starts to accelerate in $X=0.51, 0.52, 0.565$ and the last two in $X=0.566$
(c) Evolution of the scale factor $b$. Here this scale factor decreases with respect to $X$. 
(d) Evolution of the scale factor $c$. Here this scale factor is a decreasing function.
(e) Evolution of the EoS $\omega_e$.
As we can see, there are initial conditions  that allow $\omega_e$ to have values $<-1$ and the lowest value of $\omega_e$ is $-2.1$.} 
\label{figura3}
\end{figure}

In the case of $\bar{g}_4$ we find
\begin{eqnarray}
a''&=&\frac{2}{15}\frac{\Omega_{0\gamma}b_0c_0^2}{a^3bc^2}-\frac{7}{5}\frac{a'^2}{a}+\frac{1}{5}\frac{c'^2a}{c^2}\nonumber\\
 &-&\frac{4}{5}\frac{a'c'}{c}+\frac{a}{H_0^2}+ \left(\frac{1}{10c^2}-\frac{b^2}{40c^4}\right)\\ &+&\left[\frac{-3a'^2}{5a}\frac{-ac'^2}{5c^2}\frac{-6a'c'}{5c}-\frac{a}{10H_0^2 c^2}+\frac{ab^2}{40H_0^2c^4}+\frac{b_0c_0^2}{5bc^2}\left(\frac{\Omega_{0m}}{a^2}+\frac{\Omega_{0\gamma}}{a^3}\right)\right]\left(\frac{-2a'+2\frac{c'}{c}a}{3a'+2\frac{c'}{c}a}\right)\nonumber\\
b'&=&\frac{\frac{-3a'^2bc}{a}-\frac{ac'^2b}{c}-6a'c'b+\frac{1}{H_0^2}\left(-\frac{ab}{2c}+\frac{ab^3}{8c^3}\right)+\frac{b_0c_0^2}{c}\left(\frac{\Omega_{0m}}{a^2}+\frac{\Omega_{0\gamma}}{a^3}\right)}{3a'c+2c'a}\\ \nonumber\\
c''&=&\frac{3}{5}\frac{a'^2c}{a^2}-\frac{4c'^2}{5c}-\frac{9a'c'}{5a}+\frac{1}{H_0^2}\left(-\frac{2}{5c}+\frac{9b^2}{40c^3}\right)-\frac{\Omega_{0\gamma}b_0c_0^2}{5a^4bc}\\ \nonumber\\
&+&\left[\frac{-3a'^2c}{5a^2}-\frac{{c'}^2}{5c}-\frac{6a'c'}{5a}-\frac{1}{10H_0^2c}+\frac{b^2}{40H_0^2c^3}+\frac{b_0c_0^2}{5bc}\left(\frac{\Omega_{0m}}{a^3}+\frac{\Omega_{\gamma}}{a^4}\right)\right]\left(\frac{3a'c-3c'a}{3a'c+2c'a}\right)\nonumber
\end{eqnarray}
We find again initial conditions for a decelerating initial expansion that, at some point, becomes an accelerating one, we can see this in Figure \ref{figura3}b where we plot $a''$ with respect to $X$.  In Figure \ref{figura3}e we can see that $w_e$ takes negative values that depend on the initial conditions. In this case, $\omega_e$ is always negative in $z=0$ because the scale factor starts to accelerate before is equal to 1. From Figure \ref{figura3}a and Figure \ref{figura3}e we can see that, when the scale factor reaches the value of $1$ in a time approximately to the age of the universe $(X=1)$, the value of $\omega_e$ is $\approx -2.1$, but it starts accelerating in $X\approx 0.6$. 
Unlike  metric $\bar{g}_3$, there are initial conditions that allow the lowest value of $\omega_e$ to be $-2.1$ while with the $\bar{g}_3$, the lowest value of $\omega_e$ is less than $-1$.
In Figure $\ref{figura3}$ we can see that while the scale factor grows up $b$ and $c$ decrease.

\section{Discussion and Conclusions}
\label{sec:conclusions}

In this work we followed the hypothesis that the universe is, topologically, a principal fiber bundle, where the base space is the space-time. We test different Lie groups for the fiber, compact and non-compact. We calculate the Einstein field equations on the base space and study the effects of the fiber on the base space looking for those that give some interesting dynamics for the base space. We define the global EoS of the system, taking into account that this definition will be the one that the observers will measure using telescopes.
We find that a fiber of one and two dimensional Lie group does not reproduce our observed universe, it is necessarily a group of higher dimensions. With this aim we study two different 3-dimensional Lie groups as fibers, namely the group $SU(1,1)$ and the group $SU(2)$. 
We find that for the $SU(1,1)$ group there are accelerating solutions of the space-time but the values of $\omega_e$ do not correspond to the one observed in the real universe. Nevertheless, we find that for the compact group $SU(2)$ there are accelerating solutions that give a similar behavior for the EoS as the observed recently in galaxies surveys \cite{Veto}, coming to the conclusion that the accelerating behavior of the universe could be a consequence of the whole topology of the universe, more than a strange kind of matter. It remains open whether there exist a group that fits the new date and observations using this formulation and whether the fluctuations in this model are in concordance with the observed CMB and mass power spectrum. This will be reported elsewhere in the near future.

\acknowledgments
This work was partially supported by CONACyT M\'exico under grants CB-2011 No. 166212, CB-2014-01 No. 240512, Project
No. 269652 and Fronteras Project 281;
Xiuhcoatl and Abacus clusters at Cinvestav, IPN; I0101/131/07 C-234/07 of the Instituto
Avanzado de Cosmolog\'ia (IAC) collaboration (http://www.iac.edu.mx/).  M.H. acknowledge financial support from CONACyT doctoral fellowship. Works of T.M. are partially supported by Conacyt through the 
Fondo Sectorial de Investigaci\'on para la Educaci\'on, grant CB-2014-1, No. 240512

\end{document}